\documentclass[a4paper]{jpconf}
\usepackage{graphicx}

\newcommand{\be}{\begin{equation}}
\newcommand{\ee}{\end{equation}}
\newcommand{\ba}{\begin{eqnarray}}
\newcommand{\ea}{\end{eqnarray}}

\begin{document}
\begin{flushleft}
KCL-PH-TH/2010-33
\end{flushleft}
\title{Quantum Mechanical Aspects of Cell Microtubules: Science Fiction or Realistic Possibility?}

\author{Nick E. Mavromatos}

\address{CERN. Theory Division, CH-1211 Geneva 23, Switzerland \\ On leave from: King's College London, Physics Department, Strand, London WC2R 2LS, UK}

\ead{nikolaos.mavromatos@kcl.ac.uk}

\begin{abstract}
Recent experimental research with marine algae points towards quantum entanglement at ambient temperature, with correlations between essential biological units separated by distances as long as 20 Angstr\"oms. The associated decoherence times, due to environmental influences, are found to be of order 400 fs. This prompted some authors to connect such findings with the possibility of some kind of quantum computation taking place in these biological entities: within the decoherence time scales, the cell ``quantum calculates'' the optimal ``path'' along which energy and signal would be transported more efficiently. Prompted by these experimental results, in this talk I remind the audience of a related topic proposed several years ago in connection with the possible r\^ole of quantum mechanics and/or field theory on  dissipation-free energy transfer in microtubules (MT), which constitute fundamental cell substructures. The basic assumption was to view the cell MT as quantum electrodynamical cavities, providing sufficient isolation \emph{in vivo} to enable the formation of electric-dipole quantum coherent solitonic states across the tubulin dimer walls. Crucial to this, were argued to be the electromagnetic interactions of the dipole moments of the tubulin dimers with the dipole quanta in the ordered water interiors of the MT, that play the r\^ole of quantum coherent cavity modes. Quantum entanglement between tubulin dimers was argued to be possible, provided there exists sufficient isolation from other environmental cell effects. The model was based on certain ferroelectric aspects of MT. Subsequent experiments \emph{in vitro} could not confirm ferroelectricity at room temperatures, however they provided experimental measurements of the induced electric dipole moments of the MT under the influence of external electric fields. Nevertheless, this does \emph{not} demonstrate that \emph{in vivo} MT are not ferroelectric materials. More refined experiments should be done. In the talk I review the model and the associated experimental tests so far and discuss future directions, especially in view of the algae photo-experiments.

\end{abstract}

\section{Introduction: Quantum Mechanics and Biology: fiction or fact?}
\paragraph{}
\vspace{0.2cm}
It is a common perception that Quantum Mechanics (QM) pertains to the \emph{small} (\emph{micro}scopic) and \emph{cold}, whilst classical physics affects the large (\emph{macro}scopic) and complex systems, usually embedded in relatively \emph{hot environments}.

Elementary particles are from the above point of view the best arena for studying quantum effects, and this has lead to important discoveries regarding the structure of our Universe at microscopic scales, of length size less than $10^{-18}$~m~\cite{pdg}.
However, there are important examples from condensed matter physics where quantum effects manifest themselves
at relatively \emph{large distances} and/or \emph{high temperatures}. A famous example is the superconductivity phenomenon~\cite{sc}, where
formation of electron pairs exhibiting quantum coherence at macroscopic scales of order of a few thousand Angstr\"oms (\emph{i.e}. at distances three orders of magnitude larger than the atomic scale) is responsible for electric-current transport virtually dissipation free. Subsequently, high-temperature superconductors, with critical temperatures for the onset of superconductivity up to 140 K (albeit with coherence length of order of a few Angstr\"oms), have also been discovered~\cite{hightc}. In ref.~\cite{caes} the authors describe the
\emph{macroscopic}  entanglement of two samples of Cs atoms (containing more than $10^{12}$ atoms) at \emph{room
temperature}. Quite recently, it has also been demonstrated experimentally that in certain polymer chains, and under certain circumstances, one may observe quantum phenomena associated with intrachain (but not interchain) coherent electronic energy transport at \emph{room temperature}~\cite{collini}.

A natural question, therefore, which comes to one's mind is whether similar quantum phenomena may occur in biological systems, which are certainly complex, relatively large (compare to atomic physics scales) entities,
living at room temperatures. In fact this is an old question, dating back to Schroedinger~\cite{schr},  who in his famous 1944 Book entitled ``\emph{What is life}'', attempted to argue that certain aspects of living organisms, such as mutations (changes in the DNA sequence of a cell's genom or a virus), might not be explainable by classical physics but required quantum concepts, for instance quantum leaps.

Several years later, H. Fr\"ohlich~\cite{froehlich} have suggested that macroscopic quantum coherent phenomena may be responsible for dissipation-free energy and signal transfer in biological systems through coherent excitations in the microwave
region of the spectrum due to nonlinear couplings of biomolecular dipoles. The frequency with which such coherent modes are `pumped' in biological systems was conjectured to be of order
\be\label{froehlichfreq}
t_{\rm coherence~Froehlich} \sim 10^{-11}- 10^{-12}~{\rm s}~.
\ee
which is known as Fr\"ohlich's frequency.

Soon after, A.S. Davydov~\cite{davydov}, proposed that \emph{solitonic excitation states} may be responsible for dissipation-free energy transfer along the $\alpha$-helix self-trapped amide in a fashion similar to superconductivity: there are two kinds of excitations in the $\alpha$-helix: deformational oscillations in the $\alpha$-helix lattice, giving rise to quantized excitations (``phonons''), and internal amide excitations. The resulting non-linear coupling between these two types of excitations is a Davydov soliton, which traps the vibrational energy of the $\alpha$-helix and thus prevents its distortion (solitons are classical field theory configurations with finite energy).

In a rather different approach, F. Popp~\cite{popp} suggested that studies of the statistics of counts of photons in ultra-weak bioluminescence in the visible region of the spectrum point towards the existence of a coherent component linked to the living state. There have also been attempts~\cite{swain} to link the mechanisms for quantum coherence in Biology suggested by Fr\"ohlich and Popp.

In the 1990's a suggestion on the r\^ole of quantum effects on brain functioning, and in particular on conscious perception, has been put forward by R.~Penrose and S.~Hameroff~\cite{ph}, who concentrated on the microtubules (MT)~\cite{mt} of the brain cells. In particular, they noted that one may view the tubulin protein dimer units of the MT as a quantum two-state system, in coherent superposition. The model of \cite{ph} assumes, without proof, that  sufficient environmental isolation occurs, so that the \emph{in vivo} system of MT in the brain undergoes \emph{self-collapse}, as a result of sufficient growth that allowed it to reach a particular threshold, namely a critical mass/energy, related to \emph{quantum gravity} (orchestrated reduction method). This type of collapse should be distinguished from the
standard \emph{environmental decoherence} that physical quantum systems are subjected to~\cite{zurek}. In this way, the authors of \cite{ph} argue that decoherence times of order O(1~s), which is a typical time for conscious perception, may be achieved, thereby deducing that consciousness is associated with \emph{quantum computations} in the mind.

Unfortunately, in my opinion, environmental decoherence, even for \emph{in vivo} MT, cannot be ignored. In a series of works, which I will review below~\cite{mn1,mmn}, we have developed a \emph{quantum electrodynamics cavity model} for MT, in which electromagnetic interactions between the electric dipole moments of the tubulin protein dimer units and the corresponding dipole quanta in the (thermally isolated) water interiors of the \emph{in vivo} MT, are argued to be the dominant forces, inducing environmental entanglement and eventual decoherence~\cite{zurek} in at most O($10^{-6}-10^{-7}$)~s. Such times are much shorter than the required time scale for conscious perception, but have been argued to be sufficient for \emph{dissipation-less} energy transfer and signal transduction along moderately long MT of length sizes of order $\mu$m = $10^{-6}$~m. As I will discuss below, the basic underlying mechanism is the formation of appropriate \emph{solitonic} dipole states along the protein dimer walls of the MT, which are reminiscent of the quantum coherent states in the Fr\"ohlich-Davydov approach. We have also speculated~\cite{mmn} that under sufficient environmental isolation, which however is not clear if it can be achieved in \emph{in vivo} MT systems, these coherent states may provide the basis for an operation of the MT as quantum logic and information teleporting gates. At any rate, our main concern in the above works was the search for, and modeling of, possible quantum effects in cell MT which may not be necessarily associated with conscious perception. In fact in this talk I will disentangle the latter from dissipation-free energy and signal transfer in biological matter, which I will restrict my attention to.

All the above are so far mere speculations. Until recently, there was no strong experimental evidence (if at all) to suggest that macroscopic quantum coherent phenomena might have something to do with the living matter.
There were of course consistency checks with such assumptions, as is, for instance, the work of \cite{mershin}, which, by performing experiments on the brain of \emph{Drosophila}, provided consistency checks for the cavity model of MT~\cite{mn1,mmn}, especially on its possible r\^ole  for brain memory function. Nevertheless,
despite the interesting and quite delicate nature of such experiments, one
could not extract conclusive experimental evidence for quantum aspects of the brain.

To such skepticism, I would also like to add the fact that some theoretical estimates on the environmental decoherence time in brain microtubules, performed by Tegmark~\cite{tegmark}, in a model-independent way (which, however, as I will discuss below, is probably misleading), place the relevant decoherence MT time scales in the range \begin{equation}\label{tegmark}
t_{\rm decoh~MT~estimate} \in 10^{-20} - 10^{-13}~{\rm s}~,
\end{equation}
depending on the specific environmental source. Such estimates made the author of \cite{tegmark} to suggest that there is no r\^ole of quantum physics in the functioning of the brain. For the purposes of our talk below, I notice that it is the upper limit in (\ref{tegmark}) that has been proposed in \cite{tegmark} as a conservative estimate on the
characteristic decoherence time for MT, the main source of decoherence being assumed to be the Ca$^{2\, +}$ ions in each of the 13
microtubular protofilaments. Although I would disagree with such estimates, for reasons to be explained below, nevertheless, I will point out later in the talk that, even if a MT decoherence time scale of order $10^{-13}$~s is realized in Nature, this still allows for quantum effects to play a significant r\^ole associated with optimization of efficient energy and signal transfer.

At any rate, because of such a distinct lack of experimental confirmation so far, many scientists believed that any claim on a significant r\^ole of quantum physics in biology constituted science \emph{fiction}.

\section{Recent Experimental Evidence for Biological Quantum Entanglement?}
\vspace{0.2cm}

\begin{figure}[t]
\begin{center}
  \includegraphics[width=5cm, angle=-90]{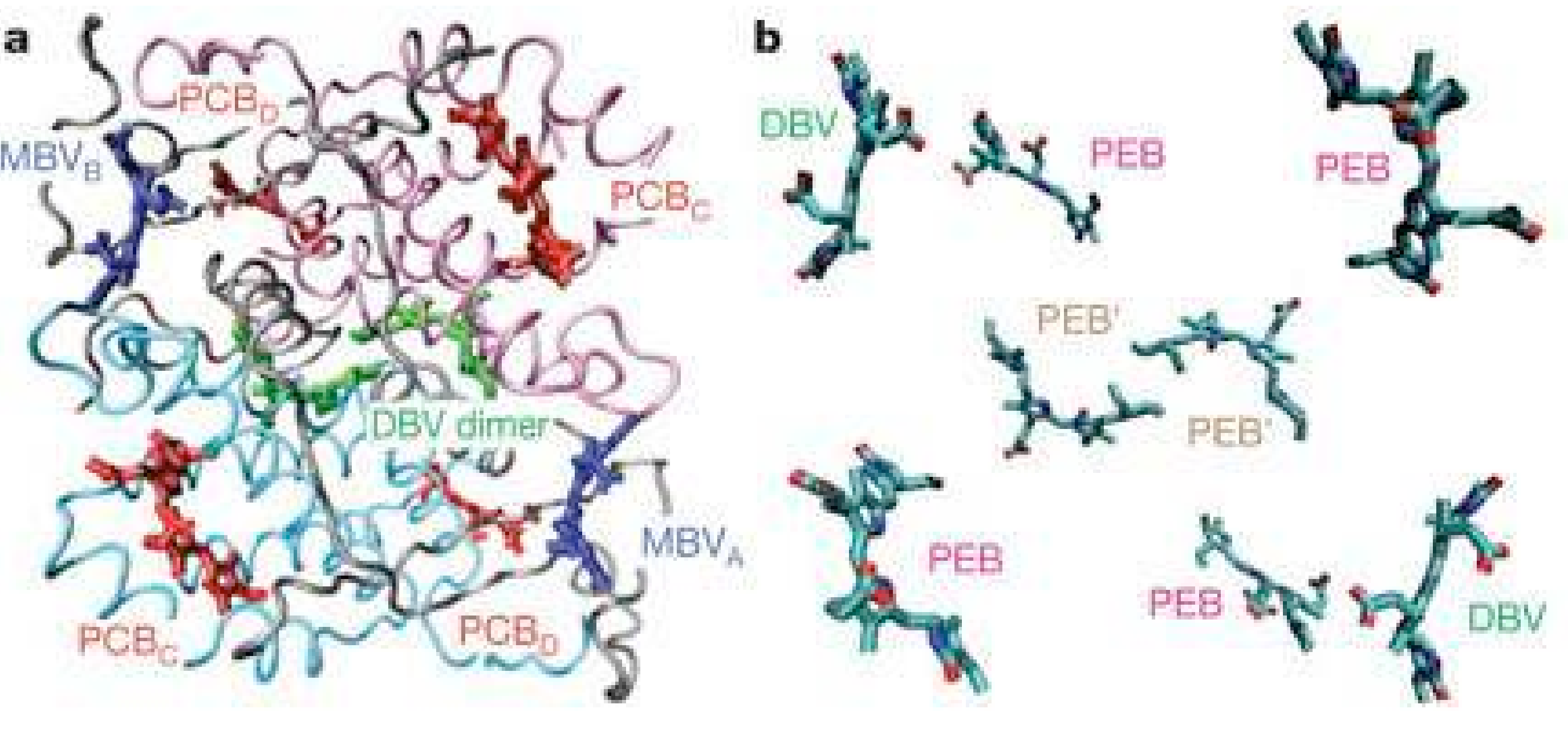}\hfill\includegraphics[width=5.5cm, angle=-90]{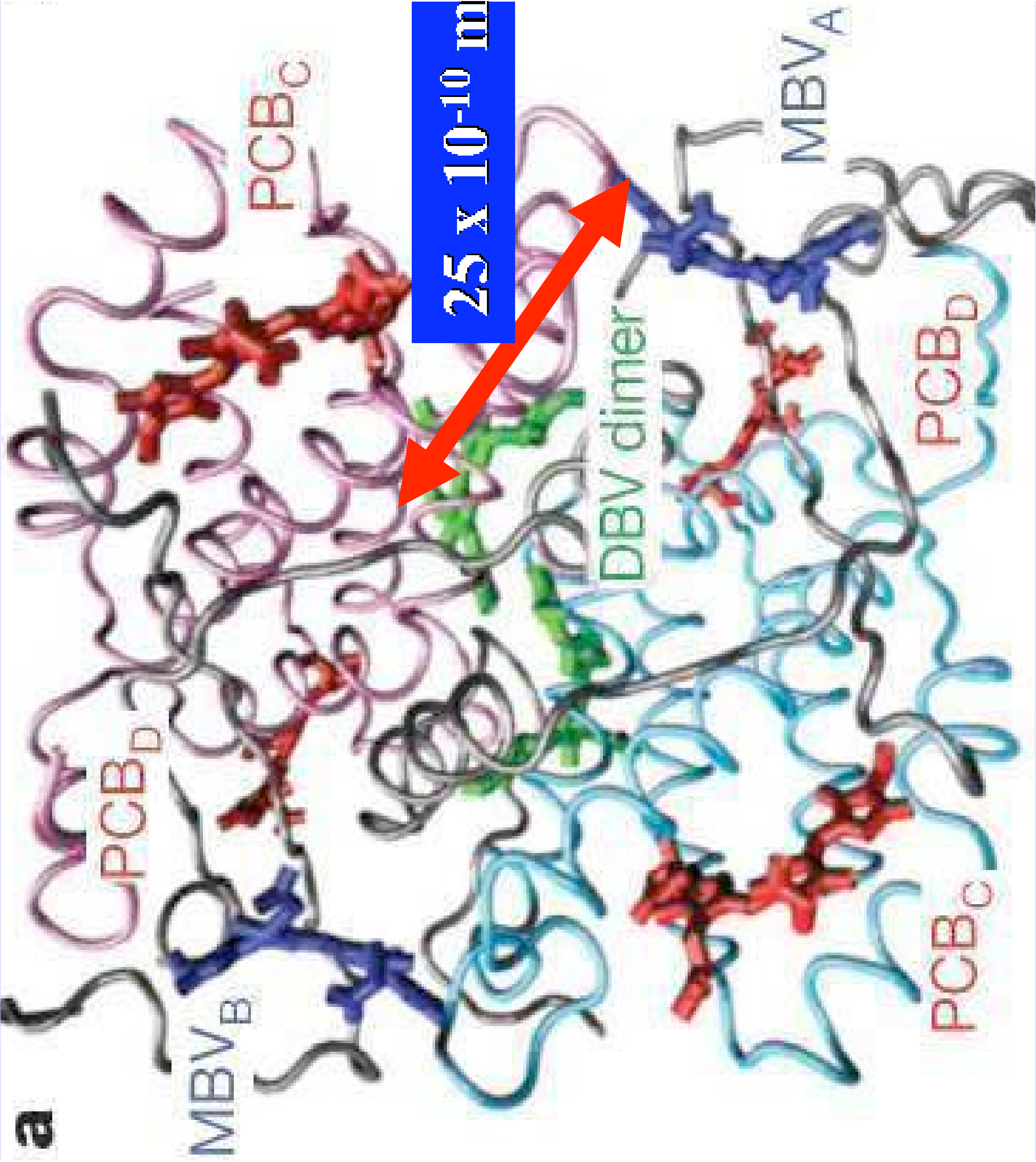}
\end{center}
\caption{\textbf{\emph{Top figures:}} \textbf{a.} Structural model of one type of  \emph{Cryptophytae Marine Algae} (CMA) protein antenna, PC645~\cite{algae}. The eight bilin molecules (Chromophores) responsible for light harvesting are indicated in various colours. \textbf{b.} The Chromophores from the structural model of the second type of CMA protein antenna studied in \cite{algae}, PE545. \emph{\textbf{Bottom figure:}} the same as in \textbf{a.} above, but with the alleged quantum-entanglement (coherent-wiring) distance of about 25 Angstr\"om between bilin molecules indicated by a red double arrow.}
\label{fig:algae}
\end{figure}
The situation concerning the experimental demonstration of a concrete r\^ole of quantum physics on basic functions of living matter started changing in 2007, when research work on photosynthesis in plants~\cite{photo} has presented rather convincing experimental evidence that light-absorbing molecules in some photosynthetic proteins capture and transfer energy according to \emph{quantum-mechanical probability laws} instead of classical laws at temperatures up to 180 K.

Even more excitingly, in the beginning of this year, compelling experimental evidence on quantum effects on living matter at ambient temperatures was provided in ref. \cite{algae}.
Using photo echo spectroscopy methods on two kinds of light-harvesting proteins, isolated appropriately from \emph{cryptophyte marine algae}, the authors of \cite{algae} have demonstrated that there exist long-lasting electronic oscillation excitations with (quantum) correlations across the 5 nm long proteins, even at \emph{room temperatures} of order 294 K.
\begin{figure}[ht]
\begin{center}
  \includegraphics[width=5cm, angle=-90]{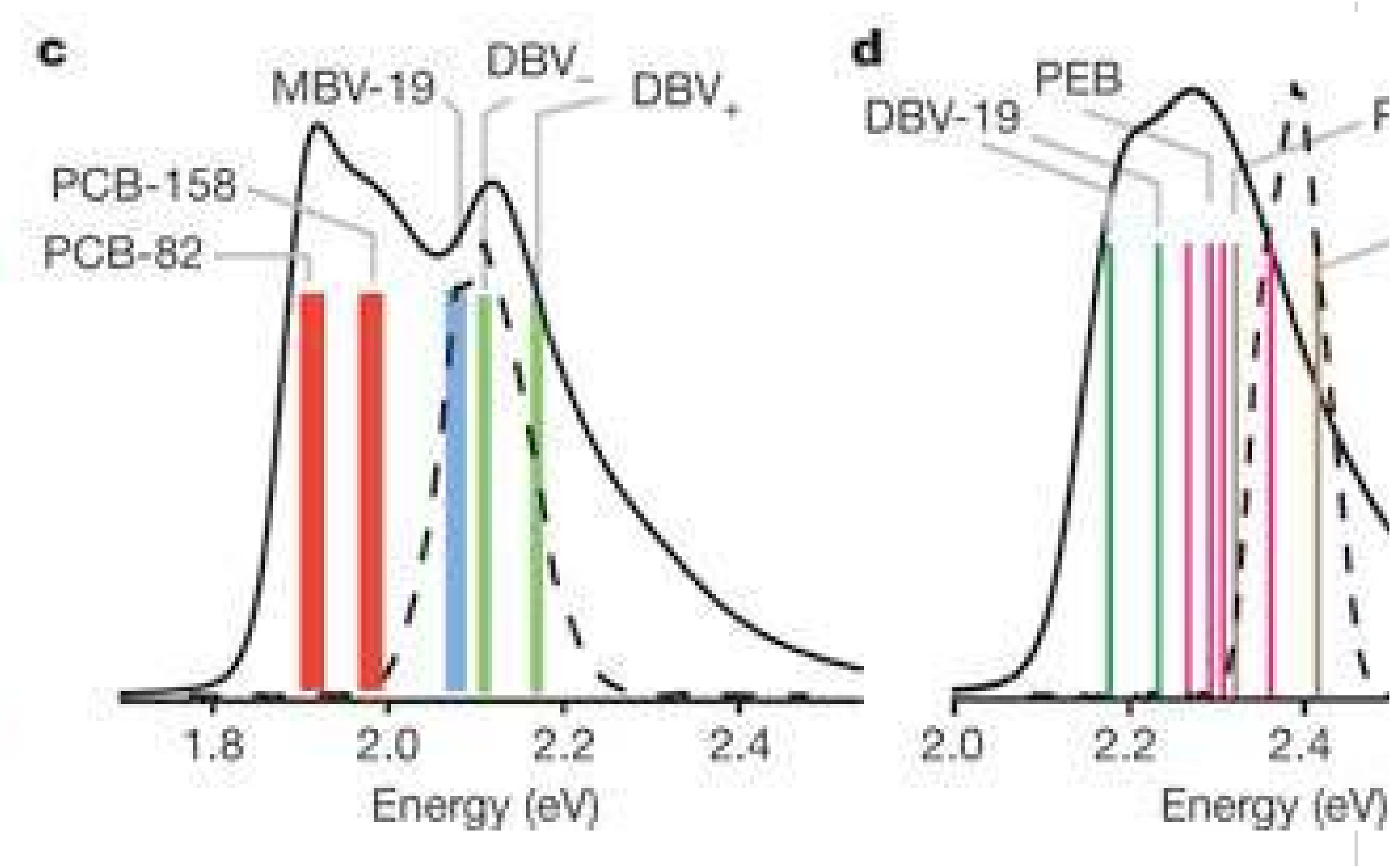}
\end{center}
\caption{\textbf{c.} The approximate electronic absorption energies of the bilin molecules indicated in fig.~\ref{fig:algae} for the PC645 protein in aqueous buffer at ambient temperatures (294 K). \textbf{d.} The same but for the protein PE545, with the same external conditions (pictures taken from \cite{algae}). The externally applied laser pulse that excites the system is indicated by a dashed line. Coloured bars denote the absorption band positions.}
\label{fig:spectrum}
\end{figure}

More specifically, there are eight light-harvesting
molecules (pigments-\emph{Chromophores}, \emph{i.e}. substances capable of changing colour when hit by light as a result of selective wavelength absorption) inside the protein antennae of marine algae (see fig.~\ref{fig:algae}). The authors of \cite{algae} studied the electronic absorption  spectrum of this complex system, and the results are indicated in fig.~\ref{fig:spectrum}.

\begin{figure}[ht]
\begin{center}
  \includegraphics[width=6cm, angle=-90]{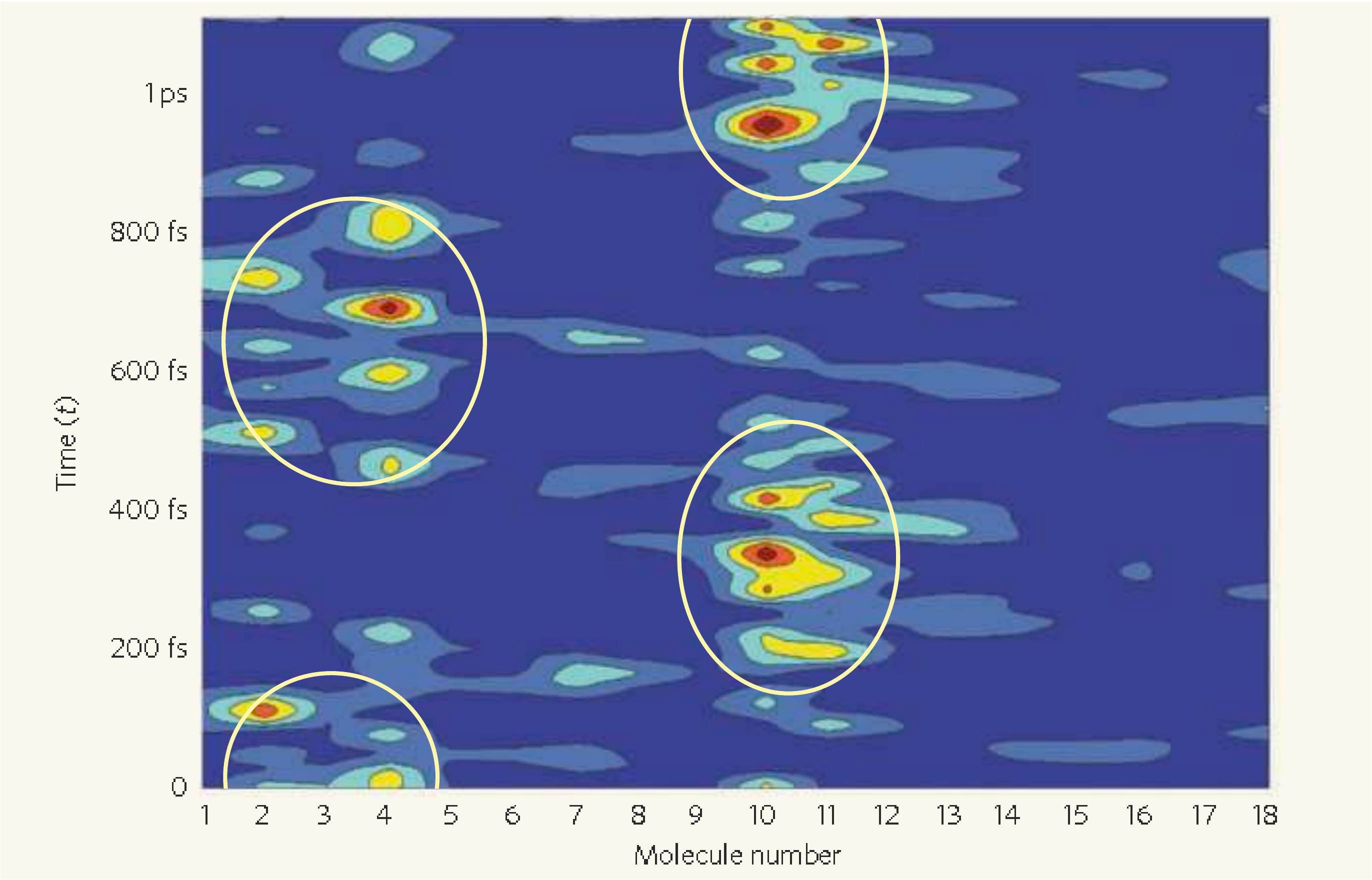}
\end{center}
\caption{Theoretical Calculations of the electronic excitation dynamics in photosynthetic light-harvesting biological complexes~\cite{photocalc}. The example is for the light-harvesting complex of the bacterium \emph{Rhodopseudomonas acidophila}, which contains eighteen pigment molecules, arranged in a circle. Upon an external photo-stimulus, such as a laser pulse, the pigments enter excited electronic states by the absorption of photons,
with coherent quantum correlations among the various pigments. The picture shows the probability of such an excitation to reside in a certain position in the complex (the dark-blue regions correspond to zero such probability, while the red-coloured regions indicate maximal probability). The vertical axis is time (in fs), while the horizontal line refers to the pigment molecules numbered 1 to 18. The white circles indicate the (small) number of molecules among which coherent oscillations of the excitation occur. In the model of \cite{photocalc} such oscillations have a half period of 350 fs. Similar oscillatory dynamics is observed in the algae complexes in \cite{algae} (\emph{c.f}. fig.~\ref{fig:echo}).}
\label{fig:photocalc}
\end{figure}

\begin{figure}[h]
\begin{center}
  \includegraphics[width=4.5cm, angle=-90]{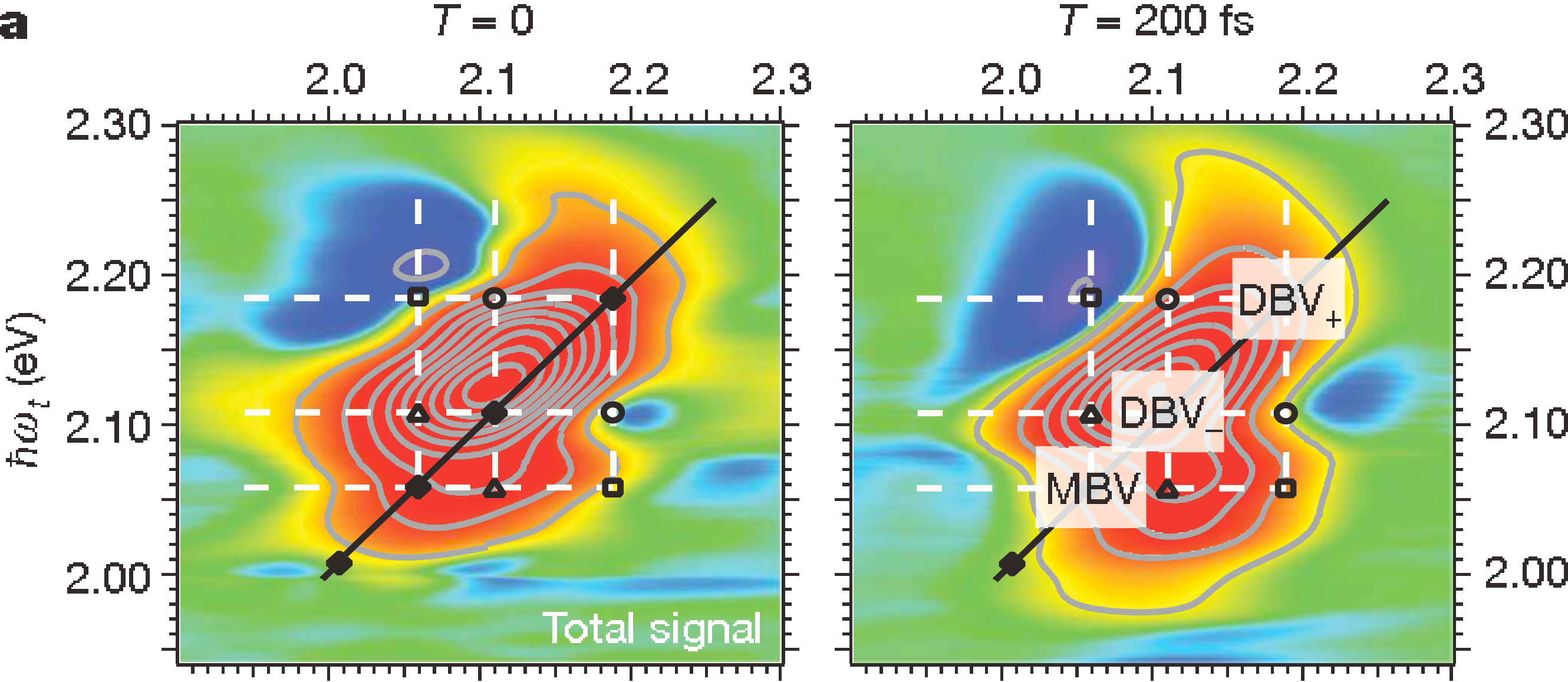}\hfill ~~~~~ \includegraphics[width=5cm, angle=-90]{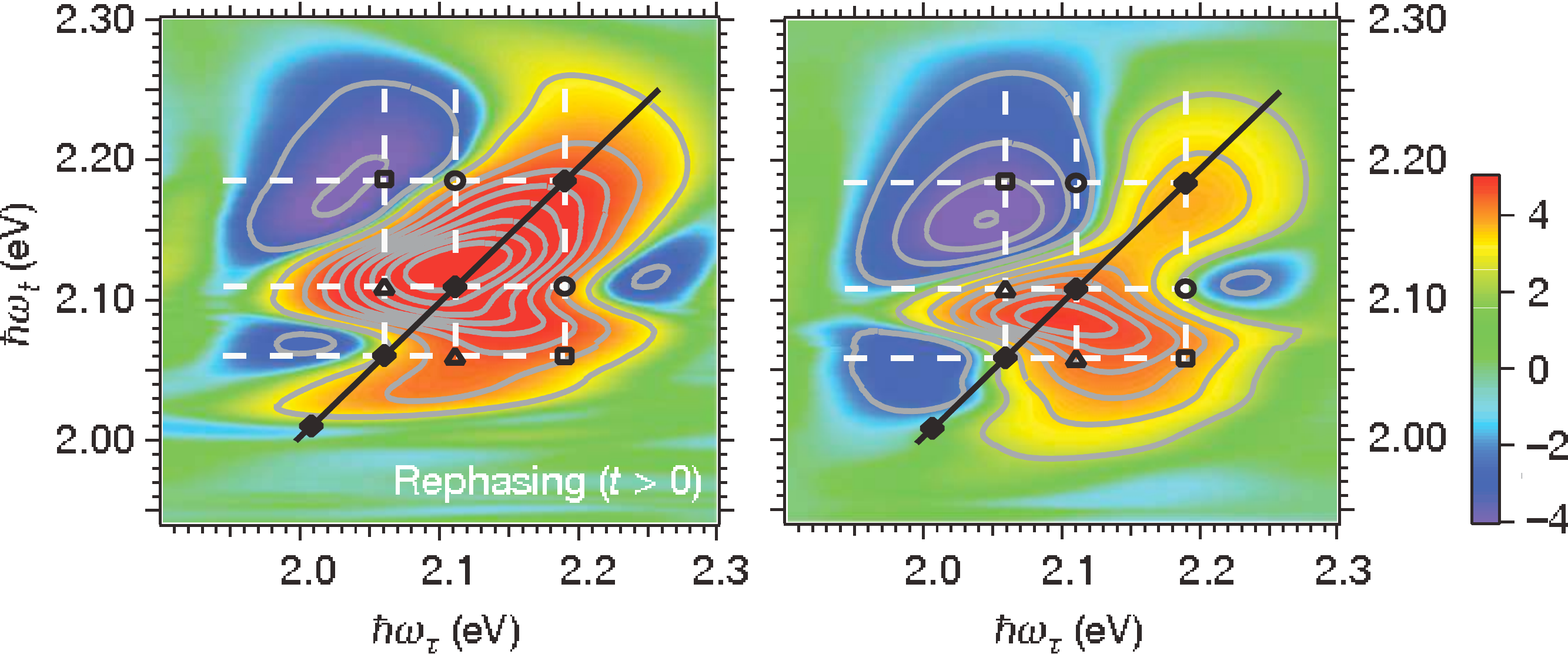}
  \vfill \includegraphics[width=5cm, angle=-90]{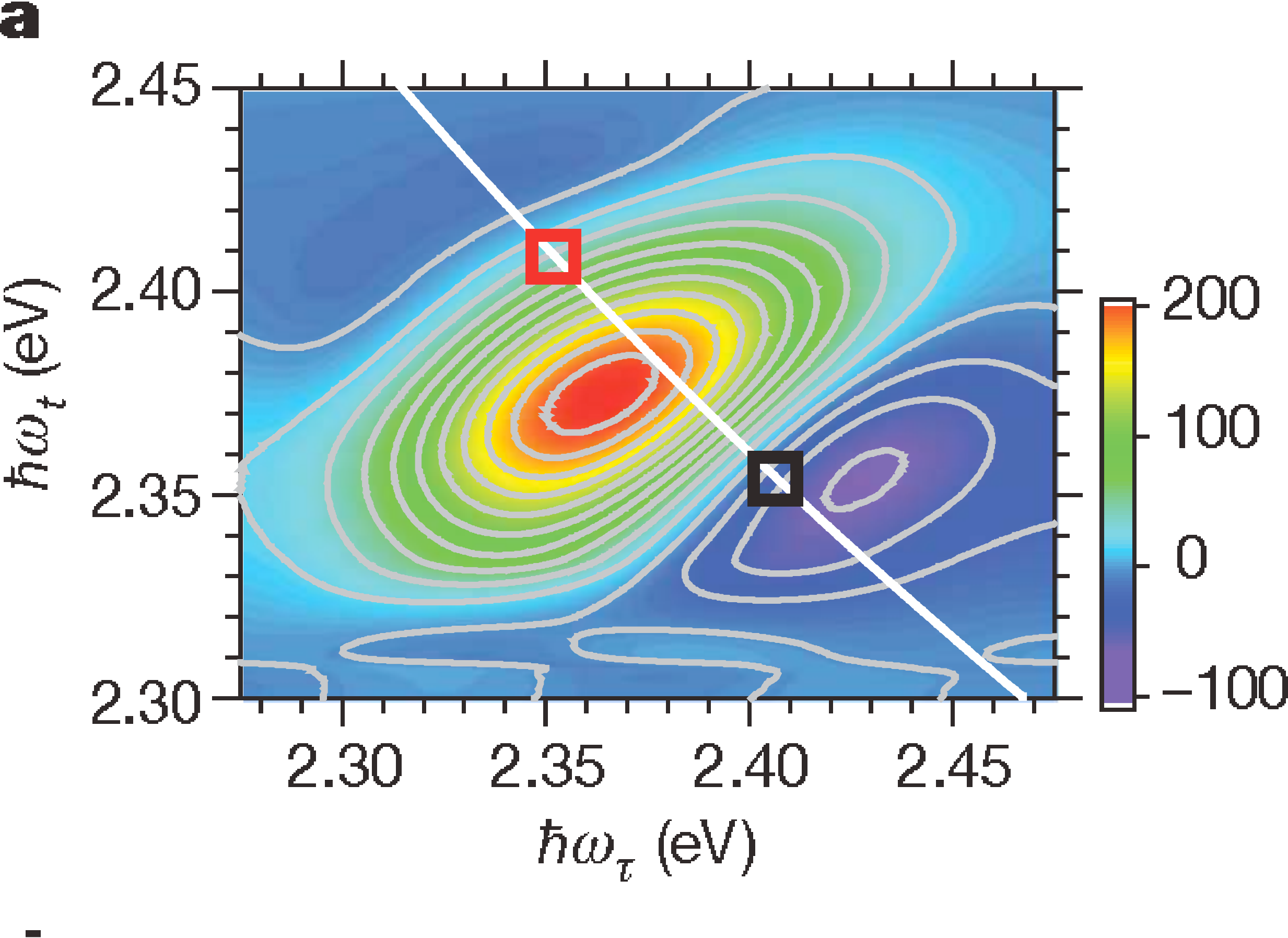} \hfill
\includegraphics[width=5cm, angle=-90]{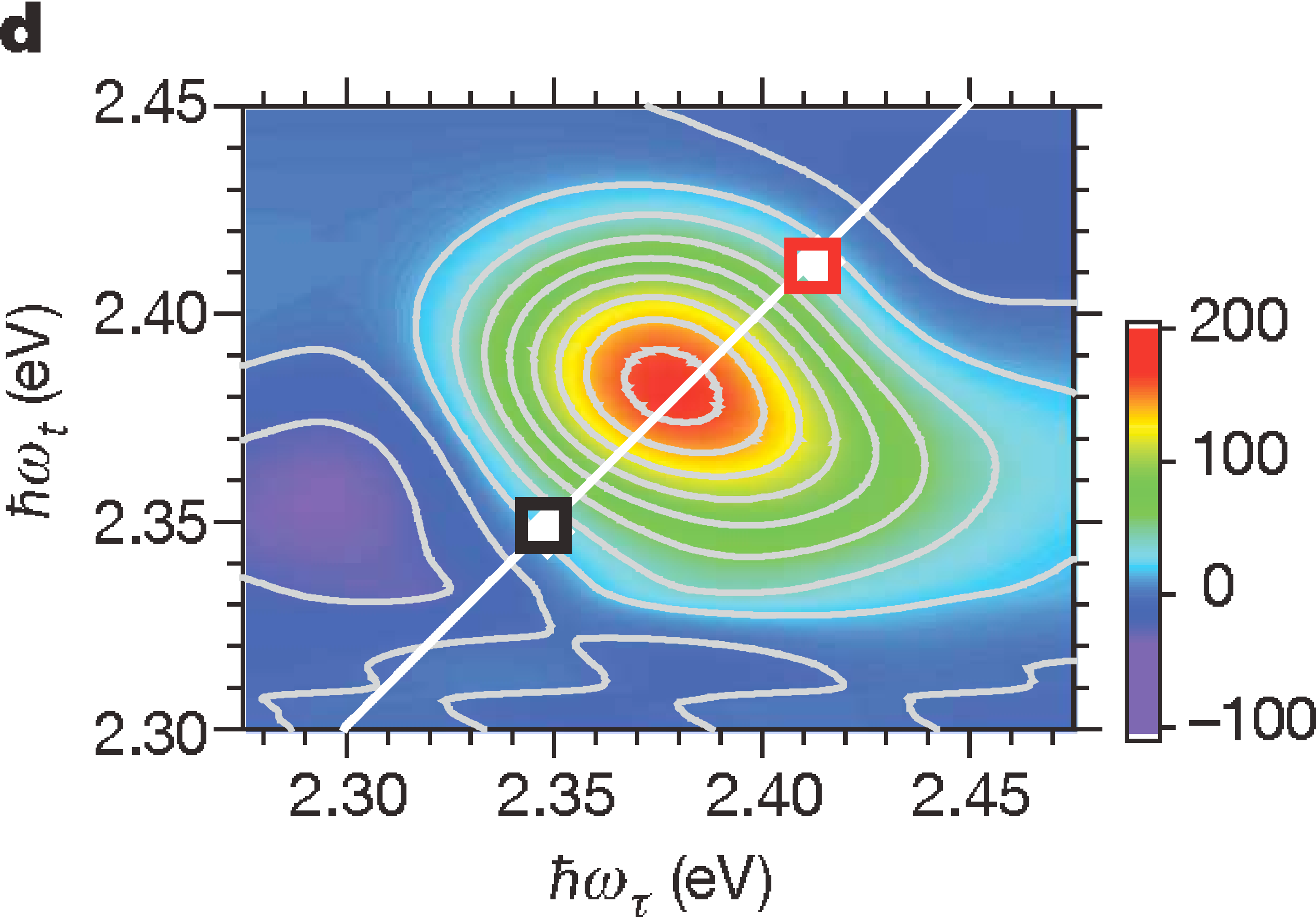}
\end{center}
\caption{The two-dimensional photon echo (2DPE) data  for PC645 protein of photosynthetic marine algae (upper picture) and for PE545 protein (lower picture).
In the upper picture \textbf{a.}, the right column shows the data for a time of 200 fs since the excitation by the external pulse. The 2DPE data show the signal intensity in an arcsinh scale plotted as a function of the coherence frequency $\omega_\tau$ and emission frequency $\omega_t$. The lower picture shows data for times 100 fs, left picture \textbf{a.} involves re-phasing of the real signal, while in picture \textbf{d.} such a re-phasing has not been performed (pictures taken from \cite{algae}).}
\label{fig:echo}
\end{figure}

In the experiments, a laser pulse (indicated by a dashed line in fig.~\ref{fig:spectrum}) of about 25 fs duration is applied to the biological entities, exciting a coherent superposition (in the form of a wave packet) of the protein antenna's vibrational-electronic eigenstates (the relevant absorption bands are indicated by coloured bars in fig.~\ref{fig:spectrum}). The relevant theory, pertaining to the quantum evolution of a system of coupled bilin molecules with such initial conditions, predicts that the relevant excitation subsequently oscillates in time
between the positions at which the excitation is localized, with distinct correlations and anti-correlations in phase and amplitude (\emph{c.f}. fig.~\ref{fig:photocalc}). Such coherent oscillations last until the natural eigenstates are restored due to \emph{decoherence}, as a consequence of environmental entanglement~\cite{zurek}. The experimental results of \cite{algae} confirmed such a behaviour (\emph{c.f}. fig.~\ref{fig:echo}), thereby indicating a quantum superposition of the electronic structure of the bilin molecule dimer DBV
(dihydrobiliverdin) for some time, which in the experiments was found to be relatively long, for a room temperature system, of order
\begin{equation}\label{algaedecoh}
t_{\rm decoh} = 400 ~{\rm fs} = 4 \cdot 10^{-13}~{\rm s}~.
\end{equation}

The quantum oscillations of the DVB molecules were transmitted to the other bilin molecules in nthe complex, at distances 20 Angstr\"oms apart, as if these molecules were connected by springs.
The authors of \cite{algae}, therefore, suggested that distant molecules within the photosynthetic proteins are `\emph{entangled}' together by \emph{quantum coherence} (``coherently wired'' is the used terminology) for more efficient light harvesting in marine algae. In other words, by exploiting such correlations, the biological cell `\emph{quantum calculates}' -- within the decoherence time scale (\ref{algaedecoh}) -- which is the most efficient way and path to transport energy across macroscopically large distances of order of a few nm (\emph{path optimization}).
Some authors would interpret this behaviour as a prototype of `quantum computation', although personally I believe we are rather far from rigorously demonstrating this.

There is an interesting question as to what guaranteed sufficient environmental isolation of the bilin molecules in the \emph{cryptophytae} antenna studied in \cite{algae} so as to have such relatively long decoherence times (\ref{algaedecoh}) at room temperatures. The authors speculate that this might be due to the fact that, unlike most of photosynthetic pigments in Nature, the eight bilin molecules in this case are covalently bound to the proteins of the \emph{cryptophytae} antenna complex.

These are, in my opinion, quite exciting experimental results that, for the first time, provide concrete evidence for quantum entanglement over relatively large distances in living matter at ambient temperature, and suggest a rather non-trivial r\^ole of quantum physics in path optimization for energy and information transport.
Given that I have worked in the past on such issues regarding Cell Microtubules (MT), it is rather natural to revisit the pertinent theoretical models, in light of the findings of ref.~\cite{algae}. It is the point of this talk, therefore, to first review the theoretical models of \cite{mn1,mmn} suggesting quantum coherent properties of MT, discuss the associated predictions as far as the scale of quantum entanglement and decoherence times are concerned, and attempt a comparison with the data of \cite{algae}.

Caution should be exercised here. Algae light-harvesting antennae and MT are entirely different biological entities. Nevertheless, they are both highly complex protein bio-structures, and the fact that quantum effects may play an essential r\^ole in energy transfer in algae at room temperature, as seems to be indicated by the recent experiments, might be a strong indication that similar coherent effects also characterize energy and signal transfer in MT \emph{in vivo}, as conjectured in \cite{mn1}. At this point, to avoid possible misunderstandings, I would like to stress once more that even if this turns out to be true,
it may have no implications for conscious perception or in general brain functioning~\cite{tus}, although, of course,  such exciting possibilities cannot be excluded.

The structure of the remainder of the talk is as follows: in the next section \ref{sec:2}, I review the basic features of the MT cavity model and discuss its predictions, especially in the light of the recent findings of \cite{algae}. In section \ref{sec:3}, I discuss some experimental tests of the model, especially as far as ferroelectric properties are concerned, which unfortunately are not conclusive. I also discuss avenues for future experiments that could confirm some other aspects of the model, including a possible extension of the photo-experiments of \cite{algae} to MT complexes. Finally, section \ref{sec:4} contains our conclusions.

\section{Cavity Model for Microtubules (MT) revisited: Quantum Coherence and Dissipation-Free Energy transfer in Biological Cells \label{sec:2}}
\begin{figure}[ht]
\begin{center}
  \includegraphics[width=6cm]{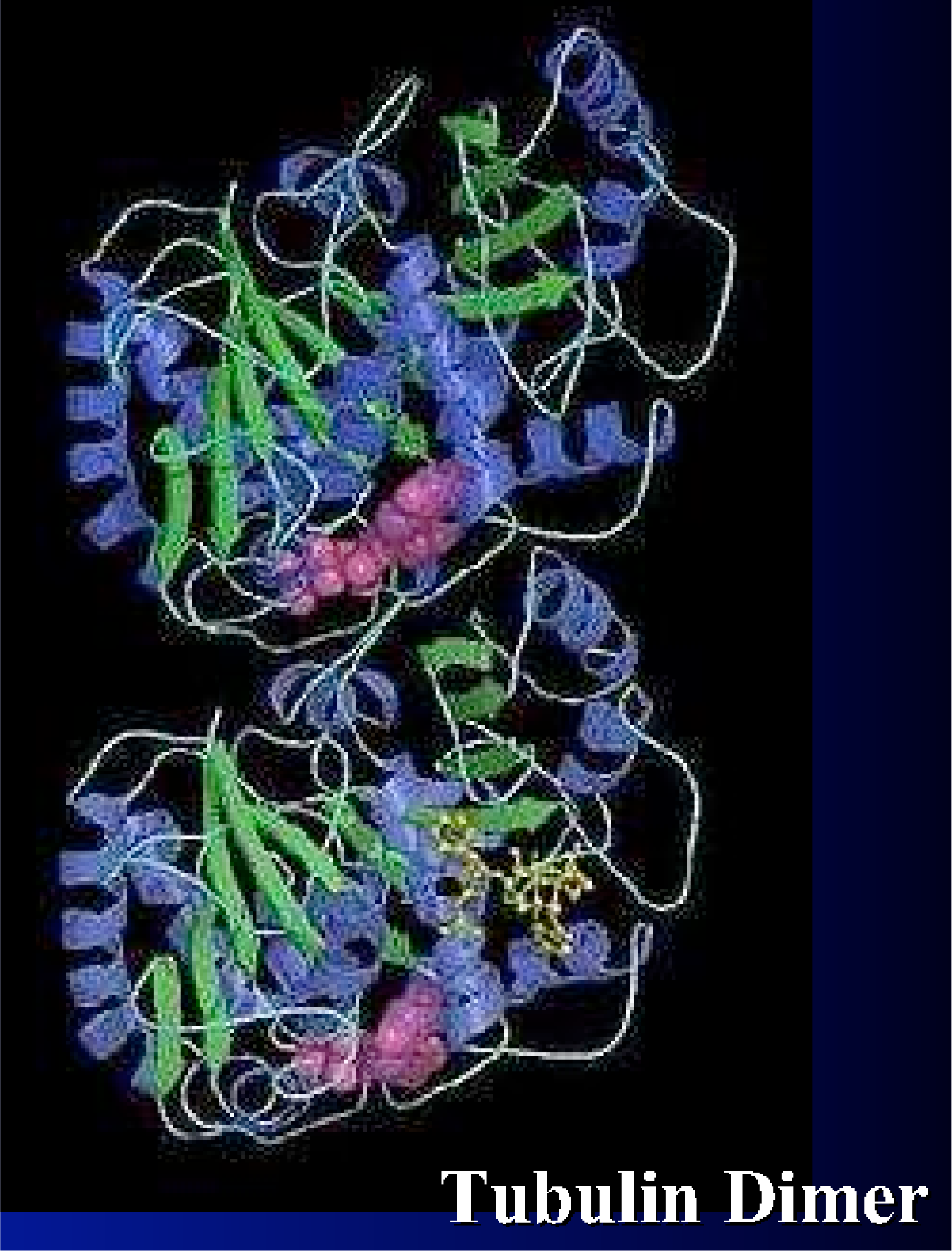}\hfill \includegraphics[width=7cm]{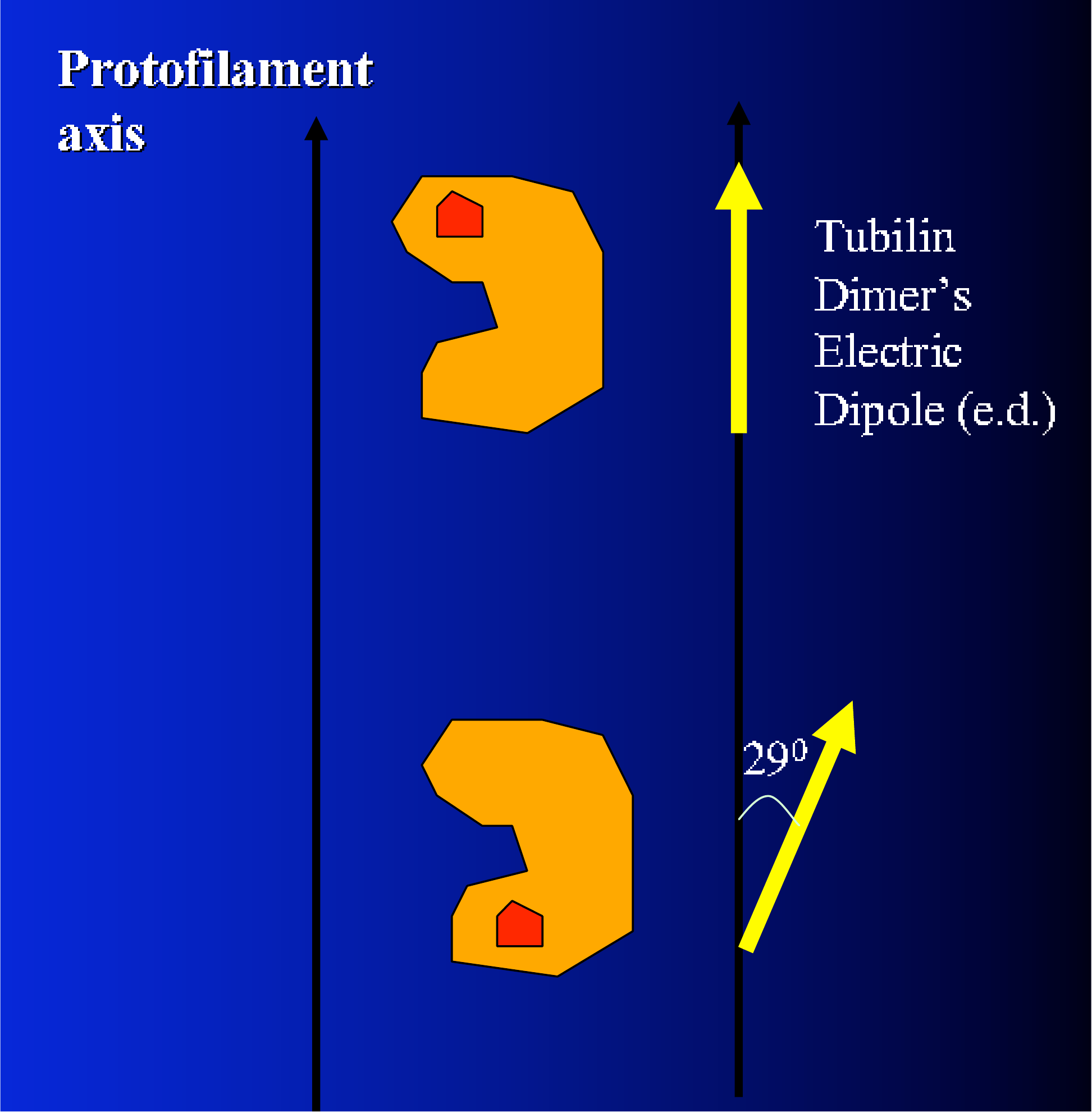}
\end{center}
\caption{\emph{\textbf{Left Picture}}: Electron Microscopy chartography of tubulin dimer at 3.5 Angstr\"om resolution~\cite{downing}. \emph{\textbf{Right Picture:}} Schematic view of the two conformations $\alpha$, $\beta$ and the position (relative to the MT axis) of the electric dipole moment. The two conformations arise from the position of the unpaired electric charge (red polygon) relative to the two hydrophobic pockets of the tubulin dimer~\cite{mt}.
In simplified models, the two monomer conformations differ by a relative displacement of the monomer's electric-dipole orientation by an angle of about 29$^0$ relative to the protofilament axis. More complicated geometries for the permanent electric dipole moment of the tubulin dimer may occur in realistic systems, according to detailed simulations~\cite{tusedm,mershinedm}, in which the bulk of it lies in a direction perpendicular to the main symmetry axis of a MT. }
\label{fig:dimer}
\end{figure}

\vspace{0.2cm}
Microtubules (MT)~\cite{mt} are paracrystalline \emph{cytoskeletal} structures that constitute the fundamental scaffolding of the cell. They play a fundamental r\^ole in the cell \emph{mitosis} and are also believed to play an important r\^ole
in the transfer of electric signals
and, more general, of energy
in the cell.
\begin{figure}[ht]
\begin{center}
\includegraphics[width=4cm]{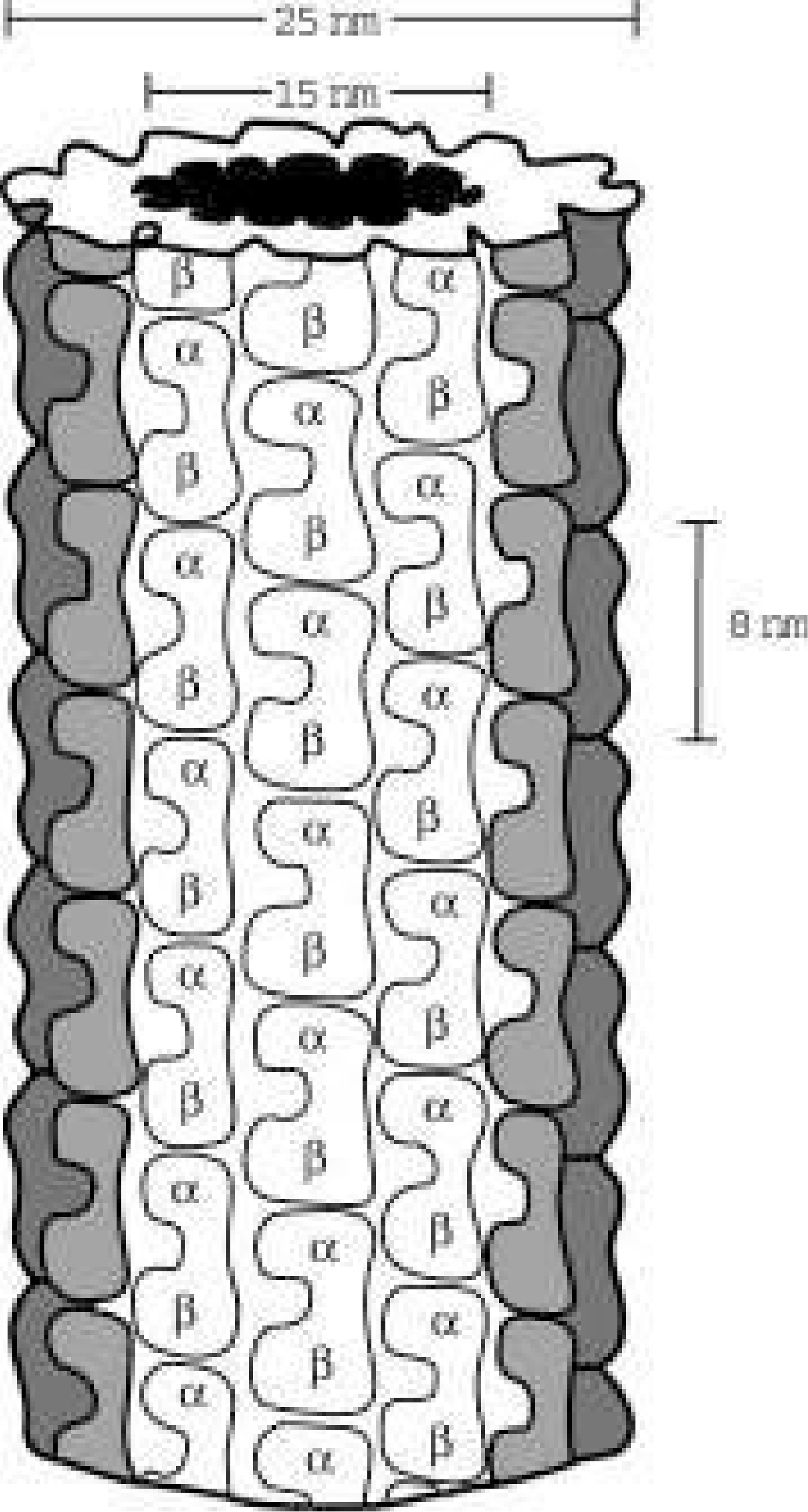}
  \hspace{2cm}\includegraphics[width=6cm]{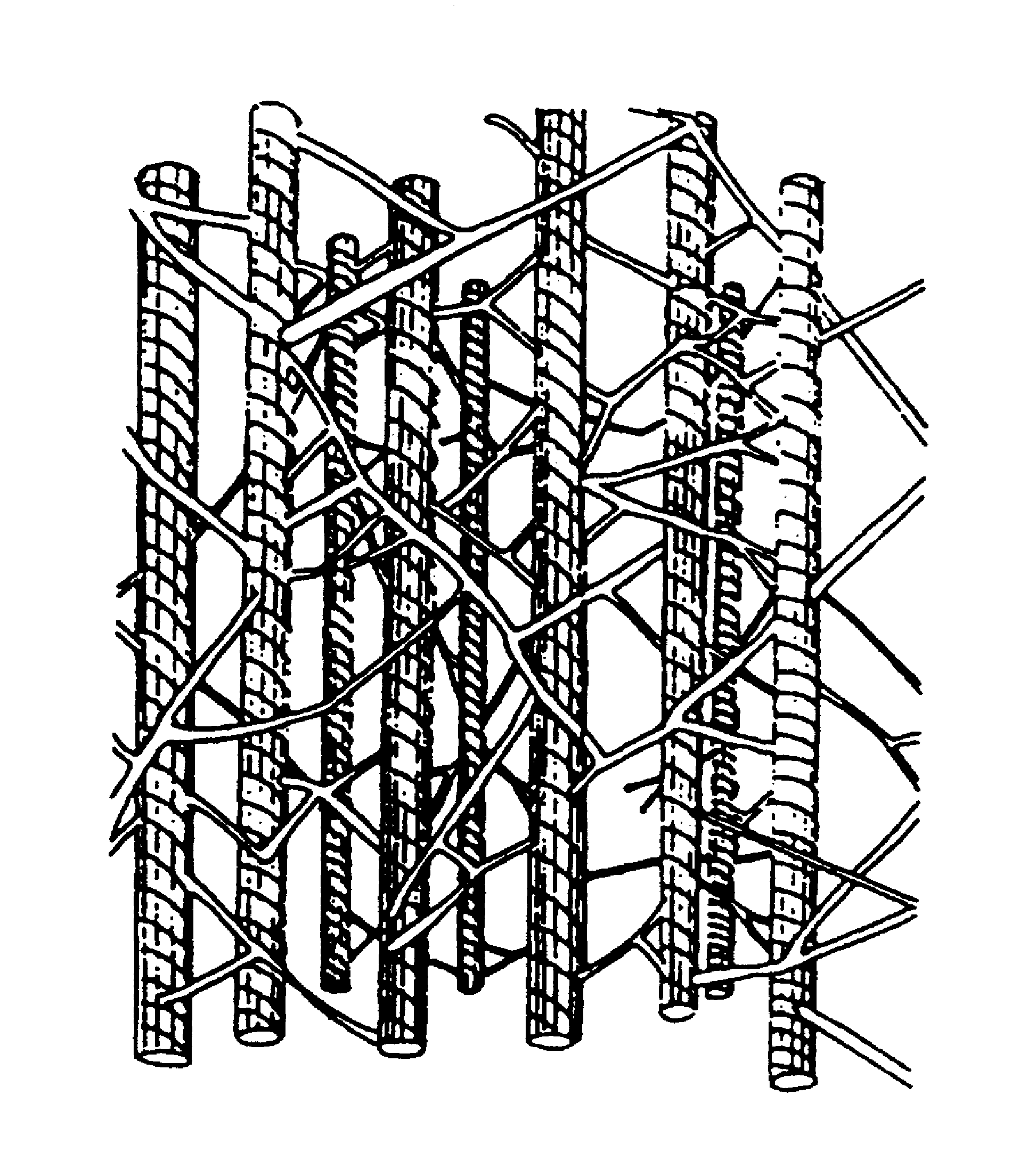}
\end{center}
\caption{\emph{\textbf{Left Picture}}: A Microtubule arrangement in the cell. The exterior walls consist of tubulin protein dimers which are arranged in 13 protofilaments (vertical chain-like structures, parallel to the main axis of MT). The interiors are full of ordered water molecules. \emph{\textbf{Right Picture:}} A network of Microtubules in a cell. The subtance connecting the various MTs together are the Microtubule Associated Proteins (MAP) (figures from ref.~\cite{sataric}).}
\label{fig:MT}
\end{figure}
They are cylindrical structures (\emph{c.f}. fig.~\ref{fig:MT}) with external cross section diameter of about 25 nm and internal diameter 15 nm. A moderately long MT may have a length of the order of a few $\mu$m = 10$^{-6}$~m.
Their exterior walls consist of tubulin protein units (\emph{c.f}. fig.~\ref{fig:dimer}). The tubulin protein dipers are characterized by two hydrophobic pockets, of length 4 nm = $ 4 \cdot 10^{-9}$~m each (the total length of a dimer being about 8 nm), and they come in two conformations, $\alpha -$ and $\beta -$ tubulin, depending on the position of the unpaired charge of 18 e relative to the pockets.

The tubulin also has an electric dipole moment.
A complete electron microscope chartography of the tubulin protein dimer is available today at 3.5 Angstr\"om resolution~\cite{downing}. This allows for theoretical modelling and computer calculations of the electric dipole moment of the dimers as well as of the entire MT~\cite{tusedm,mershinedm}.
Current simulations have shown that the bulk of the tubulin's electric dipole moment lies on an axis \emph{perpendicular} to the protofilament axis of the MT and only a fifth of the total tubulin dipole moment lies parallel to it (see also discussion in section \ref{sec:3}). However, for the purposes of constructing a (rather simplified) model of MT dynamics~\cite{sataric, mn1} that captures the essential features of dissipation-free energy and information transfer, it suffices to observe that
the two conformations of the tubulin dimer differ by a relative angle (of about 29$^0$) relative to the protofilament axis in the monomers orientation. This will have implications on the electric dipole moment of the monomer, as indicated in fig.~\ref{fig:dimer}. In this simplified picture, one ignores the components of the electric dipole perpendicular to the  protofilament's axis, and concentrates rather on a description of the
array of the dipole oscillators along the MT protofilaments by a single effective degree of freedom, namely the projection, $u_n$, on the MT cylinder's axis of the displacement of the $n$-th tubulin monomer from its equilibrium position. The strong uniaxial dielectric anisotropy
of the MT supports this picture, which enables one to view the MT as one-space dimensional crystals.

As we shall discuss below, this rather simplified geometry captures essential features of the MT, insofar as soliton formation and dissipation-free energy transfer are concerned. It is understood, though, that microscopic detailed simulations of the complete MT, which recently started becoming available~\cite{recentMTstr}, should eventually be used in order to improve the theoretical modelling of MT dynamics~\cite{sataric,mn1} and allow for more accurate studies of their possible quantum entanglement aspects.

\subsection{Classical Solitons in MT and dissipation-free energy transfer}
 \vspace{0.2cm}
 Based on such ingredients, the authors of \cite{sataric} have attempted to discuss a \emph{classical physics} model for dissipationless energy transfer across a MT, by conjecturing ferroelectric properties for MTs at room temperatures, and thus describing the essential dynamics by means of a lattice ferroelectric-ferrodistortive one-spatial dimensional chain model~\cite{collins}.

In the approach of \cite{sataric}, as mentioned previously, the relevant degree of freedom was the displacement vector $\vec{u}$ arising from the projections of the electric dipole moments of the $\alpha -$ and $\beta -$ tubulin conformations onto the protofilament axis. As a result of the inter and intra-chain interactions, this vector may well be approximated by a \emph{continuously} interpolating variable, at a position $x$ along a MT protofilament, which at time $t$ has a value
\begin{equation}\label{dipole}
    u(x, t)
\end{equation}
The time dependence is associated with the dipole oscillations in the dimers.

The relevant continuum Hamiltonian, obtained from the appropriate Lattice model ~\cite{collins} reads then~\cite{sataric}:
\begin{equation}
H = k R_0^2 (\partial_x u)^2 - M (\partial_t u)^2 - \frac{A}{2} u^2 + \frac{B}{4}u^4 + q E u~, \quad A = -|{\rm const}|\, (T-T_c)~,
\label{contham}
\end{equation}
where the critical temperature $T_c$ for the on-set of ferroelectric order is assumed at room temperatures.
$k$ is a stiffness parameter, $R_0$ is the equilibrium lattice spacing between adjacent dimers and
the term linear in $u$ is due to the influence of an external electric field of intensity $\vec{E}$, assumed parallel to the protofilament  axis ($x$ axis for concreteness in the model we are discussing);
$q=36$e is the mobile charge (the reader should recall that in MT there is an unpaired charge 18 e in each dimer conformation ) and $M$ is the characteristic mass of the tubulin dimers. The simple double-well $u^4$ non-linear potential terms in (\ref{contham}) have been assumed in \cite{sataric} to describe inter-protofilament interactions.  In \cite{mn1,mmn} we have generalized such potential terms to arbitrary polynomial terms $V(u)$ of a certain degree.

The presence of ordered water in the MT interior (fig.~\ref{fig:MT}) is approximated in this approach by the \emph{addition} to the \emph{equations of motion} derived from the Hamiltonian (\ref{contham}) of a friction term, linear in the time derivative of $u$, with a phenomenological coefficient $\gamma$:
\begin{equation}\label{eqom}
M\frac{\partial^2 u(x,t)}{\partial t^2} - k R_0^2 \frac{\partial^2 u(x,t)}{\partial x^2}
- A u + B u^3 + \gamma \frac{\partial u(x.t)}{\partial t} - q E = 0.
\end{equation}
As is well known from mechanics, when friction is present, the standard lagrangian formalism breaks down, unless one enhances the degrees of freedom of the system to include the dynamics of the environment, in order to provide a microscopic dynamical description of the friction. We shall come to this point later on. For the moment, we take into account the presence of the ordered water environment and its effects on the dimers merely by the above-mentioned friction term in eq.~(\ref{eqom}). This friction should be viewed as an environmental effect, which
however does not lead to energy dissipation, as a result of the
formation of non-trivial
solitonic structures in the
ground-state of the system
and the non-zero constant
force due to the electric field.
This is a well known result, directly relevant to
energy transfer in biological systems \cite{lal}.
\begin{figure}[t]
\begin{center}
  \includegraphics[width=8cm, angle=-90]{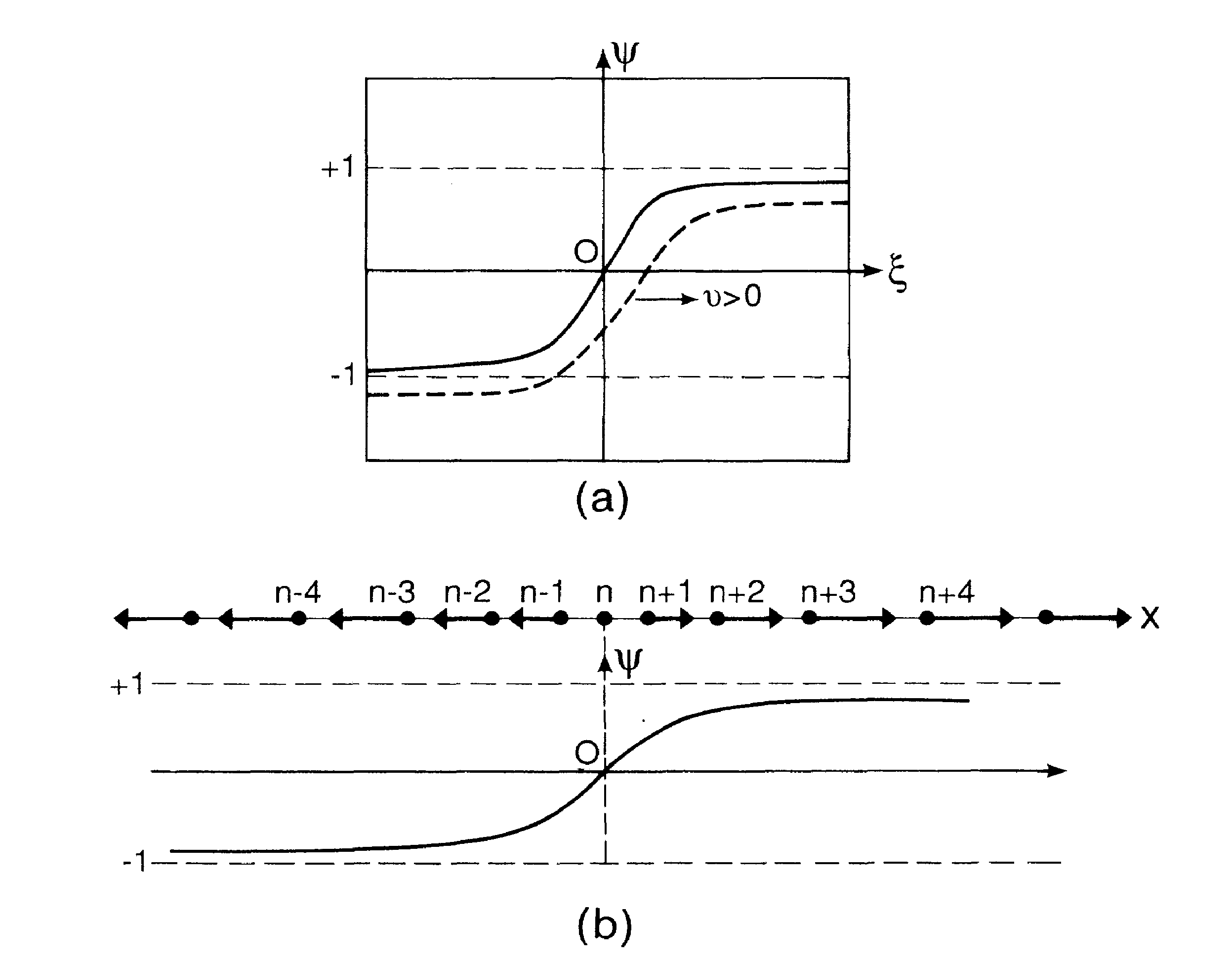}
\end{center}
\caption{The kink soliton solution (\ref{ten}) of friction equation (\ref{eleven}), along a protofilament ($x$-) axis of a MT, which is a travelling wave with velocity $v$ along the $x$ direction. The points $n$ in the lower picture denote dimer positions in an one-dimensional lattice chain, representing the protofilament tubulin dimers (from ref.~\cite{sataric}).}
\label{fig:kink}
\end{figure}
Indeed, equation (\ref{eqom}) admits as a unique bounded solution a
\emph{kink} \emph{soliton}, acquiring the form of a travelling wave~\cite{sataric} (see fig.~\ref{fig:kink})~\footnote{In \cite{mn1,mmn} we have generalized the double-well $u^4$ potential terms to arbitrary polynomial terms $V(u)$ of a certain degree.
In the mathematical literature~\cite{otinowski} there has been a
classification of solutions of friction equations (\ref{eleven}) with generalized potentials $V(u)$. For certain
forms of the potential~\cite{mn1,mmn} the solutions include {\it  kink
solitons} with the more general structure:
\begin{equation}\label{general}
u(x,t) \sim c_1 \left({\rm tanh}[c_2(x-v t)] + c_3 \right) \label{kink}
\end{equation}
where $c_1,c_2, c_3$ are constants depending on the parameters of
the dimer lattice model. For the double-well potential of \cite{sataric}, of course, the
soliton solutions reduce to (\ref{ten}).}:
\begin{eqnarray}\label{ten}
\psi (\xi ) &=& a + \frac{b -a}{1 + e^{\frac{b-a}{\sqrt{2}}\xi}}~, \\
 \psi \equiv  \frac{u(\xi)}{\sqrt{A/B}}~, \quad \xi &=& \alpha (x - u t)~, ~\quad \alpha \equiv \sqrt{\frac{|A|}{M(v_0^2 - v^2)}}~, \nonumber
\end{eqnarray}
with $\psi (\xi)$ satisfying the equation (primes denote derivatives with respect to $\xi$):
\ba\label{eleven}
 \psi '' + \rho \psi ' - \psi ^3 + \psi + \sigma = 0~, \qquad
\rho \equiv \gamma v [M |A|(v_0^2 - v^2)]^{-\frac{1}{2}}, \qquad
\sigma = q \sqrt{B}|A|^{-3/2}E~,
\ea
and the parameters $b,a$ and $d$ are defined as:
\be
(\psi -a )(\psi -b )(\psi -d)=\psi ^3 - \psi -
\left(\frac{q \sqrt{B} }{|A|^{3/2}} E\right)
\label{twelve}
\ee
The quantity
\begin{equation}
v_0 \equiv \sqrt{k/M} R_0~.
\label{sound}
\end{equation}
is the ``sound'' velocity, which is
of order 1 km/s for the system at hand~\cite{sataric}.

The kink (\ref{ten}) propagates along the protofilament axis
with fixed velocity
\be
    v=v_0 [1 + \frac{2\gamma^2}{9d^2M|A|}]^{-\frac{1}{2}}
\label{13}
\ee
This velocity depends on the strength of the electric
field $E$ through the dependence of $d$ on $E$ via (\ref{twelve}).
Notice that, due to \emph{friction}, $v \ne v_0$, and this is essential
for a non-trivial second derivative term in (\ref{eleven}), necessary
for wave propagation. For realistic biological systems $v \simeq 2$ m/s (although under certain circumstances it may even be of order
20~m/s)~\cite{sataric}).

With a velocity of this order,
the travelling waves
of kink-like excitations of the displacement field
$u(\xi )$ transfer energy
along a moderately long microtubule
of length $L =10^{-6} m$ in about
\be
t_T = 5 \times 10^{-7} {\rm s}~.
\label{transfer}
\ee

The total energy of the kink solution (\ref{ten}) is
easily calculated to be:
\be
   E = \frac{1}{R_0} \int _{-\infty}^{+\infty} dx H
= \frac{2\sqrt{2}}{3}\frac{A^2}{B} +  \frac{\sqrt{2}}{3}
k \frac{A}{B} + \frac{1}{2} M^{*} v^2  \equiv \Delta + \frac{1}{2}
M^{*} v^2
\label{energy}
\ee
and is {\em conserved} in time.
The `effective' mass $M^{*}$ of the kink is given
by
\be
            M^{*} = \frac{4}{3\sqrt{2}}\frac{MA\alpha }{R_0 B}
\label{effmass}
\ee
The first term of equation (\ref{energy})
expresses the binding energy of the kink
and the second the resonant transfer energy.
In realistic
biological models the sum of these two terms, of order of $1~{\rm eV}$,
dominate over the third
term~\cite{sataric}. On the other hand,
the effective mass in (\ref{effmass}) is of order
$5 \times 10^{-27} kg$, which is about
the proton mass ($1 GeV$) (!).

This amount of energy (\ref{energy}) is then transferred across a moderately long MT in a time scale (\ref{transfer}), virtually \emph{free} from \emph{dissipation}.

The above {\it classical}
kink-like excitations (\ref{ten})
have been discussed in \cite{sataric} in connection
with physical mechanisms associated with
the hydrolysis of GTP (Guanosine-ThreePhosphate)
tubulin dimers to GDP (Guanosine-DiPhosphate) ones.

However, because the two forms of tubulins correspond
to different conformations $\alpha$ and $\beta$, it is conceivable to speculate that
the \emph{quantum mechanical oscillations} between
these two forms of tubulin dimers
might be associated with a \emph{quantum version}
of kink-like excitations in the MT network, in which the solution (\ref{ten}) is viewed as a \emph{macroscopic coherent state}. In such a picture, this state would be itself the result of
\emph{incomplete environmental decoherence}~\cite{zurek}.
Indeed, it is known that certain quantum systems, with specific Hamiltonian interactions with the environment,
undergo \emph{incomplete} localization of their quantum states, in the sense that localization stops before it is complete. In such cases, a quantum coherent (`\emph{pointer}') state is formed as a result of decoherence~\cite{pointer}. This is argued to be the case in the so-called cavity model of MT proposed in \cite{mn1}, which we now come to review.

\subsection{Quantization of the Soliton Solutions}
\vspace{0.2cm}

Mathematically speaking, a semiclassical quantization of solitonic states of the form (\ref{ten}) (or more generally (\ref{general})), has been
considered in \cite{mn1} in a way independent of any microscopic model (for the water-induced friction).

To this end, one assumes the existence of a canonical
second quantized formalism for the $(1+1)$-dimensional
scalar field $u(x,t)$, based on creation and
annihilation operators
$a^\dagger _k$,
$a_k$.
One then constructs a squeezed
vacuum state~\cite{tdva}
\be
     |\Psi (t) > = N(t) e^{T(t)} |0>
\qquad ; \qquad T(t) = \frac{1}{2} \int \int
dx dy u(x) \Omega (x,y,t) u(y)
\label{sq1}
\ee
where $|0>$ is the ordinary vacuum state
annihilated by $a_k$,
and
$N(t)$ is a normalization factor to be determined.
$\Omega (x,y,t)$ is a complex function,
which can be split
in real and imaginary parts
as
\ba
  \Omega (x,y,t) &=& \frac{1}{2}[G_0^{-1}(x,y) - G^{-1}(x,y,t)]
+ 2i\Pi (x,y,t) \nonumber \\
G_0 (x,y) &=& <0|u(x)u(y)|0>
\label{sq2}
\ea
The squeezed coherent state for this system can be then defined
as~\cite{tdva}
\be
 |\Phi (t) > \equiv e^{iS(t)}|\Psi (t)> \qquad ; \qquad
S(t) = \int_{-\infty}^{+\infty} dx[D(x,t)u(x) - C(x,t)\pi (x)]
\label{sq3}
\ee
with $\pi (x)$ the momentum conjugate to $u(x)$, and
$D(x,t)$, $C(x,t)$ real functions.
With respect to this state, $\Pi (x,t)$ can
be considered as a momentum canonically conjugate
to $G(x,y,t)$ in the following sense
\be
         <\Phi (t) | -i\frac{\delta}{\delta \Pi (x,y,t)}
|\Phi (t)> = - G(x,y,t)
\label{sq4}
\ee
The quantity $G(x,y,t)$ represents the\emph{ modified
boson field} around the soliton. From the expression (\ref{sq3}),
one may also identify $C(x,t)$ with the dynamical field representing solitonic excitations,
which in our case are the quantum-corrected solitons of the dipole configurations of the tubulin dimers in the MT wall,
\begin{equation}\label{cuq}
u_q (x,t) = C(x,t)~.
\end{equation}

To determine the functions $C$, $D$ and $\Omega$ one applies
the \emph{Time-Dependent Variational Approach}
(TDVA)~\cite{tdva} according to which
\be
\delta \int _{t_1}^{t_2} dt <\Phi (t) |(
i\partial _t - H) |\Phi (t) >  = 0
\label{sq5}
\ee
where $H$ is the canonical Hamiltonian of the system.
This leads to a canonical set of  (quantum) Hamilton
equations
\ba
{\dot D}(x,t)&=&-\frac{\delta {\cal H}}{\delta
C(x,t)} \qquad  {\dot C}(x,t) = \frac{\delta {\cal H}}
{\delta D (x,t)} \nonumber \\
{\dot G}(x,y,t) &=& \frac{\delta {\cal H}}{\delta \Pi (x,y,t)}
\qquad {\dot \Pi}(x,y,t) =\frac{\delta {\cal H}}{\delta
G(x,y,t)}
\label{sq6}
\ea
where the quantum energy functional
${\cal H}$ is given by\cite{tdva}
\be
  {\cal H} \equiv <\Phi (t) | H | \Phi (t) >= \int _{-\infty}
^{\infty} dx {\cal E} (x)
\label{sq7}
\ee
with
\ba
{\cal E} (x) = \frac{1}{2} D^2 (x,t) &+& \frac{1}{2} (
\partial _x  C(x,t))^2  + {\cal M}^{(0)} [C(x,t)] + \nonumber \\
+\frac{1}{8}<x|G^{-1}(t)|y> &+& 2 <x|\Pi (t) G(t) \Pi (t) |y>
+\frac{1}{2}lim_{x \rightarrow y} \nabla _x\nabla _y <x| G(t) |y>
 - \nonumber \\
-\frac{1}{8}<x|G_0^{-1}|y> &-& \frac{1}{2}lim _{x \rightarrow y}
\nabla _x \nabla _y <x|G_0 (t)|y>~.
\label{sq8}
\ea
Above we used the following operator notation in coordinate
representation
$A (x,y,t) \equiv <x|A(t)|y>$, and
\be
M^{(n)} = e^{\frac{1}{2}(G(x,x,t)-G_0(x,x))\frac{\partial ^2}
{\partial z^2}} U^{(n)}(z) |_{z=C(x,t)}
\qquad ; \qquad U^{(n)} \equiv d^n U/d z^n
\label{sq9}
\ee
The function $U $ denotes the potential of the original soliton
Hamiltonian, $H$.
Notice that the quantum energy functional is conserved
in time, despite the various time dependencies
of the quantum fluctuations. This is a consequence
of the canonical form (\ref{sq6}) of the Hamilton equations.

Performing the functional derivations in (\ref{sq6})
one obtains
\ba
 {\dot D}(x,t) &=& \frac{\partial ^2}{\partial x^2}
C(x,t) - {\cal M}^{(1)}[C(x,t)]   \nonumber \\
{\dot C}(x,t) &=& D(x,t)
\label{sq10}
\ea
which after elimination of $D(x,t)$, yields
a modified soliton equation for the (\emph{quantum corrected}) field
$u_{q}(x,t) = C(x,t)$ (\emph{c.f}. (\ref{cuq}))~\cite{tdva}
\be
    \partial ^2 _t u_q(x.t) - \partial _x ^2 u_q(x,t)
+ {\cal M}^{(1)} [u_q(x,t)] = 0 \label{22c} \ee
with the notation
$$ M^{(n)} = e^{\frac{1}{2}(G(x,x,t)-G_0(x,x))\frac{\partial ^2}
{\partial z^2}} U^{(n)}(z) |_{z=u_q(x,t)}~, \quad {\rm and} \quad  U^{(n)} \equiv
d^n U/d z^n~. $$
The quantities $M^{(n)}$ carry information
about the quantum corrections. For the kink soliton (\ref{kink})
the quantum corrections (\ref{22c}) have been calculated
explicitly in ref. \cite{tdva}, thereby providing us with a
\emph{concrete example} of a large-scale quantum coherent \emph{squeezed} state.
The whole scheme may be thought of as
a \emph{mean-field-approach} to quantum corrections to the soliton
solutions.

Having established these facts, it is interesting to attempt to formulate a microscopic model for MT, leading to the above-described solitonic coherent states. Such a model has been proposed in \cite{mn1}
and elaborated further in \cite{mmn}, where its quantum information processing aspects have been discussed.
The basic ingredient of the model consists  of viewing the MT as \emph{quantum electrodynamical cavities}.
We next proceed to review briefly the model's basic features.

\subsection{The Quantum Cavity Model of MT \label{sec:mtcavity}}
\vspace{0.2cm}

According to the Quantum Cavity Model of MT, the two conformations $\alpha$ and $\beta$, which, as mentioned above, differ by the positions of the unpaired electrons relative to the two hydrophobic pockets of the dimer (fig.~\ref{fig:dimer}), are viewed as two different states of a two-state quantum system, excitable by an external stimulus. For instance, applying an external pulse, or an electric field,
excites a quantum superposition of these two states, and there are quantum oscillations between the two dimer configurations, which damp out after a finite time, as a consequence of environmental decoherence. This aspect is shared of course with the starting point of the Hemeroff-Penrose model for quantum MT~\cite{ph}. However the
mechanisms underlying the formation of the quantum coherent states and their eventual environmental decoherence are entirely different, as we now proceed to discuss.

In this scenario, Tubulin is viewed as a typical {\it two-state}
quantum mechanical system, where the dimers couple to
conformational changes with $10^{-9}-10^{-11} {\rm s}$
transitions, corresponding to an angular frequency $     \omega
\sim{\cal O}( 10^{10}) -{\cal O}(10^{12})~{\rm Hz}$. In \cite{mn1} we assumed the upper bound of this frequency range to
represent (in order of magnitude) the characteristic frequency of
the dimers, viewed as a two-state quantum-mechanical system: \be
     \omega _0 \sim {\cal O}(10^{12})~{\rm Hz}
\label{frequency2} \ee

In the quantum Cavity Model, the MT are viewed as \emph{quantum electrodynamical cavities}. In fact the \emph{thermally isolated cavity regions} are thin interior regions of thickness of order of a few Angstr\"oms \emph{near} the \emph{dimer} walls, in which the electric dipole-dipole interactions between the dimers and the ordered water molecules overcome thermal losses, and provide the necessary conditions for quantum coherent states to be formed along the tubulin dimer walls of the MT~\footnote{At this point we would like to point out the following. It is known experimentally~\cite{sackett}, that in a thin exterior neighborhood of MT there are areas of atomic thickness,
consisting of charged ions,  which isolate the MT from thermal
losses. This means that the  electrostatic interactions overcome
thermal agitations. It seems theoretically plausible, albeit yet unverified,
that such thermally isolated  exterior areas can also operate as
{\it cavity regions}, in a  manner similar to the areas interior
to MT. At this point it is unclear whether there exist the
necessary coherent dipole quanta in the ionic areas. Further
experimental and theoretical (simulational) work needs to be done
regarding this  issue.}.

Indeed,
if we consider two electric dipole vectors
$\vec{d}_i$, $\vec{d}_j$,
at locations $i$ and $j$
at a relative distance $r_{ij}$,
one pertaining to a water molecule, and the other to a protein dimer
in the MT chain,
then the dipole-dipole interaction has the form:
\be
E_{dd} \sim -\frac{1}{4\pi \varepsilon}\frac{3 ({\hat \eta}.{\vec d_i})
({\hat \eta}.{\vec d_j})- \vec{d}_i . \vec{d}_j}{|r_{ij}|^3}
\label{dipoledipole}
\ee
where ${\hat \eta}$ is a unit vector in the direction of $\vec{r}_{ij}$,
and
$\varepsilon $ is the dielectric constant of the medium.

First of all we note that, since each dimer has a mobile
charge~\cite{mt}: $q=18 \times 2e$, $e$ the electron charge,
one may {\it estimate} the electric dipole moment of the dimer roughly as
\be
d_{\rm dimer} \sim  36 \times \frac{\varepsilon_0}{\varepsilon} \times
1.6 \times 10^{-19}
\times 4 \times 10^{-9}~{\rm C}\,{\rm m} \sim 3 \times 10^{-18}~ {\rm C} \, {\rm m} = 90~{\rm Debye}~.
\label{dipoledimer}
\ee
where we
used the fact that a typical distance for the estimate
of the electric dipole
moment for the `atomic' transition between the $\alpha,\beta$
conformations is of ${\cal O}(4~{\rm nm})$, \emph{i.e}. of order of the
distance between the two hydrophobic dimer pockets.
We also took account of the fact that, as a result of the
water environment, the electric charge of the dimers appears
to be screened by the relative
dielectric constant of the water, $\varepsilon/\varepsilon_0 \sim 80$.
We note, however, that the biological environment of the unpaired
electric charges in the dimer may lead to
further suppression of $d_{dimer}$  in (\ref{dipoledimer})~\footnote{The reader is referred to section \ref{sec:3} below for further discussion on detailed simulations of the permanent electric dipole moment of the tubulin dimer. The order of magnitude of our estimates below, however, is not affected by such detailed considerations.}.

Assuming that the medium
between the ordered-water molecules in the layer
and the MT dimers
corresponds exclusively to the tubulin protein, with
a typical value of dielectric constant $\varepsilon \sim 10$~\cite{brown},
and taking into account the generic (conjectural)
values of the electric dipole moments for tubulin dimers (\ref{dipoledimer})
and water molecules~\cite{mn1} (see discussion below eq. (\ref{orwater})), we
easily conclude that
the dipole-dipole interactions (\ref{dipoledipole})
may overcome thermal losses
at room temperatures, $\sim k_BT$,
for distances $|r_{ij}|$ of up to a few tenths of an Angstr\"om.
Notice that for such distances the respective order of the
energies is ${\cal O}(10^{-2}~{\rm eV})$.
Such thin cavities may not be sufficient to sustain
quantum coherent modes for sufficiently long times necessary for dissipation-less energy
transfer along the MT.

However,
isolation from thermal losses could be assisted enormously by
the existence of a {\it ferroelectric} transition below some critical
temperature $T_c$, for the system of protein dimers in MT. In the models of \cite{sataric} and ours~\cite{mn1}
this is precisely what happens, with the critical temperature for the onset of ferroelectricity near ambient temperature

Ferroelectricity implies an
effective dielectric ``constant'' $\varepsilon (\omega) < 1$ in
(\ref{dipoledipole}); in such a case, these interactions can overcome thermal losses at room
temperatures for up to {\it a few} Angstr\"oms.
An additional possibility would be that for this range of frequencies
a {\it negative} (dynamical) dielectric constant arises.
This
would mean
that the dimer walls become {\it opaque} for the modes
in a certain range of frequencies~\cite{mn1} below some critical value,
thereby providing concrete support to the idea of MT behaving as isolated `cavities', by trapping such modes.
As explained in \cite{mn1}, such frequencies
occur naturally within the range of
frequencies of our model.

Inside such thin interior cavity regions,
there are \emph{quantum coherent modes}, which in \cite{mn1} have been argued to be the \emph{dipole quanta}, conjectured to exist in water in ref.~\cite{vitiello}, and discussed in this conference.
These coherent modes arise from the interaction of the
electric dipole moments of the water molecules with the quantized
radiation of the electromagnetic field~\cite{vitiello}, which may be
self-generated in the case of MT arrangements~\cite{sataric,mn1}.
The corresponding Hamiltonian interaction terms are of the form~\cite{vitiello}
\be
   H_{ow} = \sum_{j=1}^{M} [\frac{1}{2I} L_j^2 + \vec{A} \cdot{\vec d}_{ej}]
\label{orwater}
\ee
where $\vec{A}$ is the quantized
electromagnetic field in the radiation gauge,
$M$ is the number of water molecules, $L_j$ is the
total angular momentum
vector of a single molecule,
$I$ is the associated (average) moment of inertia,
and $d_{ej}$ is the
electric dipole vector
of a single molecule, $|d_{ej}| \sim 2e \otimes d_e$,
with $d_e \sim 0.2$ Angstr\"om.
As a result of the dipole-radiation interaction in
(\ref{orwater}) coherent modes emerge, which in \cite{vitiello}
have been interpreted as arising from the quantization of the
Goldstone modes responsible for the {\it spontaneous breaking}
of the electric dipole (rotational) symmetry. Such modes are termed
`\emph{electric dipole quanta}' (EDQ). It is our view that such modes do not characterize ordinary water, but may well arise in the ordered water in the MT interior, which is a different phase of water. Such quanta play a r\^ole similar to the coherent modes of quantized electromagnetic radiation in ordinary cavities.

In our cavity model for MT~\cite{mn1} such coherent modes play the role of `\emph{cavity modes}' in the quantum
optics terminology~\cite{haroche}. These in turn interact with the dimer
structures, mainly through the unpaired electrons of the dimers,
leading to the formation of a quantum coherent solitonic state
that may extend over the entire MT, and under sufficient isolation even over the entire MT network.
As mentioned
above, such states may be identified~\cite{mn1} with semi-classical
solutions of the friction equations (\ref{22c}). In the model of \cite{mn1}, such
coherent, \emph{almost classical}, states are viewed as the result
of {\it incomplete decoherence} of the dimer system due to its
interaction/coupling with the water environment~\cite{zurek}.
Incomplete decoherence may characterize some systems, in the sense that environmentally induced decoherence stops before complete localization of the quantum state. If this happens, then such partial decoherence time scales could be identified with the time taken for the dimer solitons to form.

In \cite{mn1} estimates for the \emph{formation time} of such solitons have been given, using conformal field theory methods for the description of the dynamics of the dipoles along the (one-dimensional) protofilaments, represented as Ising spin chains.
Admittedly, these are crude approximations, and thus the so-obtained formation time estimates may not be accurate.
For completebess we mention that such formation times of the solitons along the dimer walls may be of order $10^{-10}$ s, although smaller times cannot be excluded. This is the time scale over which
solitonic coherent pointer states in the MT
dimer system are formed (`\emph{pumped}'), according to our scenario, which is not far from the originally assumed Fr\"ohlich's coherence time scale (\ref{froehlichfreq}).
\begin{figure}[t]
\begin{center}
  \includegraphics[width=7cm,angle=-90]{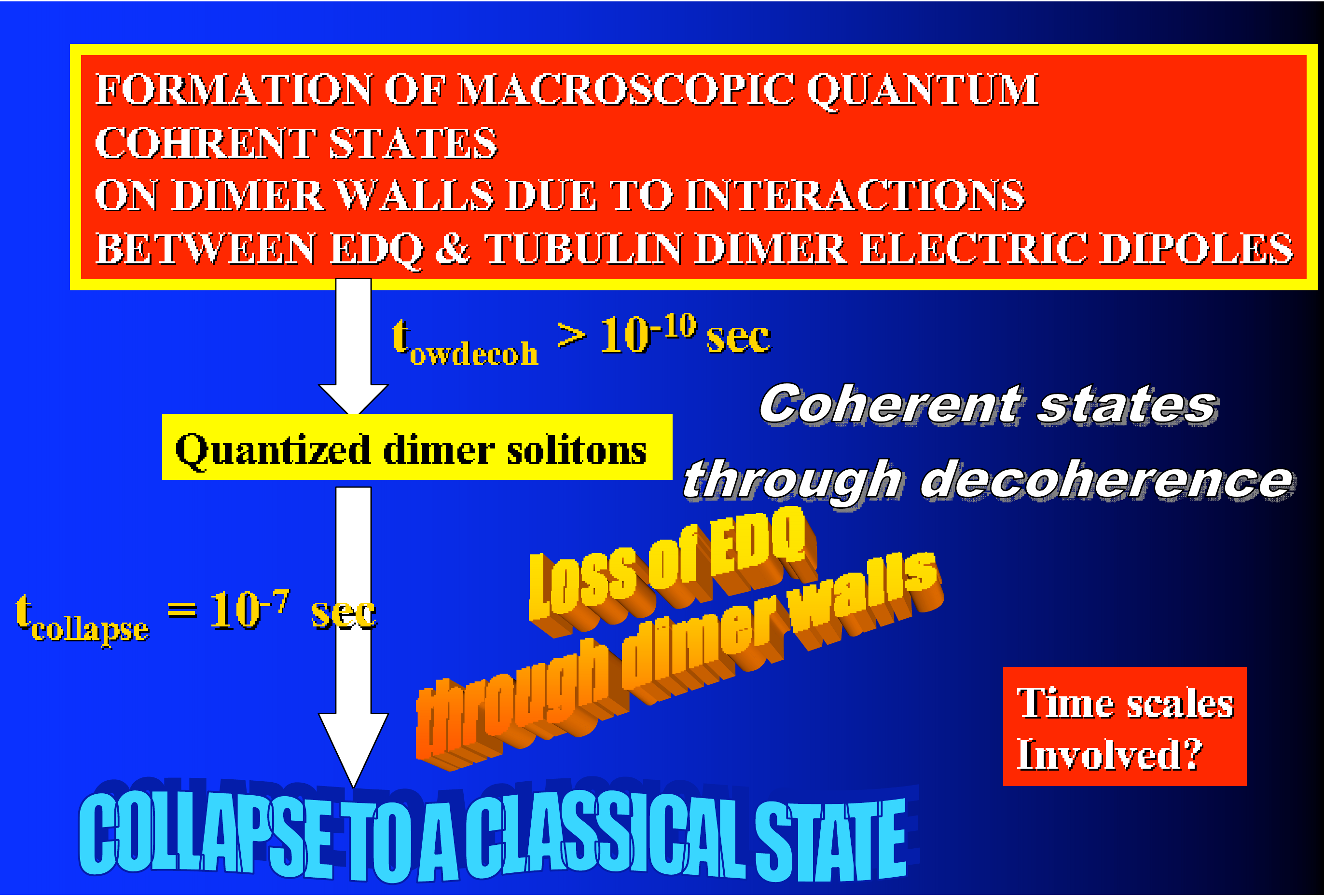}
\end{center}
\caption{In the Cavity model for MT of \cite{mn1}, there are two stages of quantum decoherence: one incomplete one, arising from the r\^ole of ordered water as an environment to the dimer system, which results in the formation of the soliton coherent state, and the second one, much longer, which results in the complete collapse of the soliton coherent state, as a result of the the loss of dipole quanta and energy through the imperfect dimer walls.
Several uncertainties as to the precise origin of the incomplete decoherence exists in the model, which may shorten significantly the first time scale, below the $O(10^{-10})$ s. For dissipationless energy transfer in MT, it is the second (complete) decoherence time scale that it is important.}
\label{fig:twoscale}
\end{figure}
Eventually, these soliton coherent states will decohere to purely classical configurations.
The decoherence scales involved in this second stage can be estimated by applying standard cavity quantum optics considerations to the MT system (see fig.~\ref{fig:twoscale}).

Indeed, the above-mentioned dimer/water coupling leads to a situation analogous to
that of atoms in interaction with coherent modes of the
electromagnetic radiation in {\it quantum optical cavities}. An important phenomenon characterizes such interactions, namely the so-called {\it Vacuum-Field Rabi Splitting} (VFRS)
effect~\cite{rabi}. VFRS appears in both the emission and
absorption spectra of atoms~\cite{agar}. For our
purposes below, we shall review the phenomenon by restricting
ourselves for definiteness to the absorption spectra case.

Consider a collection of $N$ atoms of characteristic
frequency $\omega_0$ inside an electromagnetic cavity. Injecting a pulse
of frequency $\Omega$ into the cavity causes a doublet
structure (splitting) in the absorption spectrum of the
atom-cavity system with peaks at~\cite{rabi}:
\be \Omega = \omega _0 -
\Delta/2 \pm \frac{1}{2}( \Delta ^2 + 4 N \lambda ^2 )^{1/2}
\label{rabiabs} \ee
where $\Delta = \omega_c - \omega_0 $ is the
detuning of the cavity mode, of frequency $\omega_c$, compared to
the atomic frequency. For resonant cavities the splitting occurs
with equal weights \be
  \Omega = \omega_0 \pm \lambda \sqrt{N}
\label{rabisplitting} \ee Notice here the {\it enhancement} of the
effect for multi-atom systems $N >> 1$. The quantity  $2\lambda
\sqrt{N}$ is called the `Rabi frequency'~\cite{rabi}. {}From the
emission-spectrum analysis an estimate of $\lambda$
can be inferred which involves the matrix element, ${\vec
d}$, of atomic electric dipole between the energy states of the
two-level atom~\cite{rabi}: \be
   \lambda = \frac{E_{c}{\vec d}.{\vec \epsilon}}{\hbar}
\label{dipolerabi} \ee where ${\vec \epsilon}$ is the cavity
(radiation)  mode polarisation, and \be E_{c} \sim
\left(\frac{2\pi \hbar \omega_c}{\varepsilon  V}\right)^{1/2}
\label{amplitude} \ee is the r.m.s. vacuum (electric) field
amplitude at the center of a cavity of volume $V$, and of
frequency $\omega_c$, with $\varepsilon $ the dielectric constant
of the medium inside the volume $V$. In atomic physics the VFRS
effect has been confirmed by experiments involving beams of
Rydberg atoms resonantly coupled to superconducting
cavities~\cite{rabiexp}.

In the analogy between the thin cavity regions near the dimer
walls of MT with electromagnetic cavities, the
role of atoms in this case is played by the unpaired two-state
electrons of the tubulin dimers~\cite{mn1} oscillating with a
frequency (\ref{frequency2}). To estimate the Rabi coupling
between cavity modes and dimer oscillations, one should use
(\ref{dipolerabi}) for the MT case.

In \cite{mn1} we have used some simplified models for the ordered-water
molecules, which yield a frequency of the coherent
dipole quanta (`cavity' modes) of order~\cite{mn1}: \be
     \omega_c \sim 6 \times 10^{12}~ {\rm s}^{-1}~.
\label{frequency} \ee Notably this is of the same order of magnitude as the
characteristic frequency of the dimers (\ref{frequency2}),
implying that the dominant cavity mode and the dimer system are
almost in resonance. Note that this
is a feature shared by atomic physics systems in cavities, and
thus we can apply the pertinent formalism to our system. Assuming
a relative dielectric constant of water w.r.t to that of vacuum
$\varepsilon_0$, $\varepsilon/\varepsilon_0 \sim 80$, one obtains
from (\ref{amplitude}) for the case of MT cavities: \be
    E_{c} \sim 10^{4}~{\rm V/m}
\label{eowmt} \ee Electric fields of such a magnitude can
be provided by the electromagnetic interactions of the MT dimer
chains, the latter viewed as giant electric
dipoles~\cite{sataric}. This suggests that the
coherent modes $\omega_c$, which in our scenario interact with the
unpaired electric charges of the dimers and produce the kink
solitons (\ref{ten}), (\ref{22c}) along the chains, owe their existence to  the (quantized)
electromagnetic interactions of the dimers themselves.

The Rabi coupling for an MT with ${\cal N}$ dimers, then, is estimated from
(\ref{dipolerabi}) to be of order: \ba    {\rm
Rabi~coupling~for~MT} \equiv \lambda _{MT}
= \sqrt{{\cal N}} \lambda \sim 3 \times 10^{11}~ {\rm s}^{-1}~,
\label{rabiMT} \ea which is, on average, an order of magnitude
smaller than the characteristic frequency of the dimers
(\ref{frequency2}).

In the above analysis, we have assumed that the system of tubulin dimers
interacts with a {\it single} dipole-quantum coherent mode of the
ordered water and hence we ignored dimer-dimer interactions.
More complicated cases, involving interactions
either among the dimers or of the dimers with more than one
radiation quanta, which undoubtedly occur {\it in vivo}, may affect
these estimates. Moreover, the use of more sophisticated models for the description of water, are in need.
We hope to be able to come back to such improved analyses in the near future, especially in view of the rapid recent developments on the experimental detection of quantum coherent effects in biological systems~\cite{photo,algae}.

For the time being, we note that the presence of a Rabi coupling between water molecules and dimers
provides a \emph{microscopic} description for the \emph{friction} that leads to quantum coherent solitonic states (\ref{ten}), (\ref{22c}) of the electric dipole
quanta on the tubulin dimer walls. To estimate the decoherence
time we remark that the main source of dissipation (environmental
entanglement) comes from the imperfect walls of the cavities,
which allow \emph{leakage of coherent modes and energy}.
The time scale, $T_r$, over
which a cavity-MT dissipates its energy, can be identified in our
model with the average life-time $t_L$ of a coherent-dipole
quantum state, which has been found to be~\cite{mn1}: $T_r \sim t_L
\sim  10^{-4}~{\rm s}$. This leads to a first-order-approximation estimate of the
quality factor for the MT cavities, $Q_{MT} \sim \omega_c T_r \sim
{\cal O}(10^8)$. We note, for comparison, that high-quality
cavities encountered in Rydberg atom experiments dissipate energy
in time scales of ${\cal O}(10^{-3})-{\cal O}(10^{-4})$~s, and
have $Q$'s which are comparable to $Q_{MT}$ above.

Applying standard quantum mechanics of quantum electrodynamical cavities~\cite{haroche,rabi,rabiexp},
we may express the pertinent decoherence time in terms of $T_r$, the Rabbi coupling $\lambda$
and the number of oscillation quanta $n$ included in a coherent mode of dipole quanta in the ordered water.
The result is~\cite{mn1}:
\be
    t_{collapse} = \frac{T_r}{2 n {\cal N}{\rm sin}^2\left(\frac{{\cal N}
n \lambda^2\,t}{\Delta}\right)}
\label{convtime}
\ee
The time $t$
represents the `time' of interaction of the dimer system with the
dipole quanta. A reasonable estimate~\cite{mn1} of this time scale
can be obtained by
equating it with the average
life-time of
a coherent dipole-quantum state, which, in the
super-radiance model of \cite{jibu}, used in our work,
can be estimated as~\footnote{In this model the entire interior volume of the MT is taken into account. In our cavity model, however, it is only a thin cylindrical region near the dimer walls that plays the r\^ole of the cavity. Nevertheless, this does not affect our estimates, since the density of the relevant water molecules $N_w/V$ is assumed uniform.}
\be
     t \sim \frac{c\hbar ^2V}{4\pi d_{ej}^2\epsilon N_{w}L}
\label{lifetime}
\ee
with $d_{ej}$ the electric dipole moment of a water molecule,
$L$ the length of the MT, and $N_w$ the number of water molecules
in the volume $V$ of the MT. For moderately long MT, with length $L \sim 10^{-6}~{\rm m}$ and $N_w \sim 10^8$,
a typical
value of $t$ is:
\be
         t \sim 10^{-4}~{\rm s}~.
\label{lifetime2}
\ee
On the assumption that a typical coherent mode of dipole quanta
contains an average of $n = {\cal O}(1)-{\cal O}(10)$
oscillator quanta, and for detuning between the cavity mode and the characteristic frequency
of the dimer quanta of order
$\Delta / \lambda \sim {\cal O}(10)-{\cal O}(100)$,
the analysis of ~\cite{mn1} yields the following estimate for the collapse
time of the kink coherent state of the MT dimers due to
cavity dissipation:
\be t_{collapse} \sim {\cal O}(10^{-7})-{\cal
O}(10^{-6})~{\rm s}~. \label{tdecohsoliton} \ee
This is of the same order as
the time scale (\ref{transfer}) required for energy transport
across the MT by an average kink soliton in the models of
\cite{sataric}. The result (\ref{tdecohsoliton}), then,
implies that quantum physics is relevant as far as
\emph{dissipationless} energy transfer across the MT is concerned.

In view of this specific model, we are therefore in stark \textit{disagreement}
with the conclusions of Tegmark in \cite{tegmark}, \emph{i.e}. that \emph{only}
classical physics is relevant for studying the energy and signal transfer in
biological matter. Tegmark's conclusions did not take proper account of the
possible isolation against environmental interactions, which seems to occur
inside certain regions of MT with
appropriate geometry and properties. The latter can screen decoherening effects in the immediate environment of an MT, for instance those due to neighboring ions used in \cite{tegmark} to arrive at the estimate (\ref{tegmark}) for the decoherence time of MT.

However, I must stress once again that the above estimates are very crude. We do not have a good microscopic model for
the ordered water in the MT interiors, nor we understand the detailed structure of the water-dimer couplings.
Above, we have ignored interactions among dimers, when we considered the dipole quanta-dimer coupling, assuming that
there is a single coupling between a dipole quantum and a dimer oscillator. Moreover, there is no fundamental reason why the detuning $\Delta$ should have the order assumed. All these complicate accurate estimates of the decoherence time, and the actual situation may be very different. One needs, first of all, to develop theoretical models for the ordered water, and use sophisticated numerical simulations in order to estimate the various quantum optical characteristics of the model, such as vacuum Rabi splittings, dipole electric moments \emph{etc}. This is not done as yet. We hope that such detailed modelling will be available in the near future. Thus, although we disagree with Tegmark's conclusion that quantum effects play no essential role in energy and information process in biological systems, such as the brain, one cannot exclude the possibility that decoherence time estimates as short as (\ref{tegmark}) may, after all, characterize MT. As I will argue below, though, this is still a long enough time interval to allow for an essential r\^ole of quantum physics to be played in the functioning of MT.

\subsection{MT Cavity Model Parameters Revisited in Light of Algae Experiments}
\vspace{0.2cm}
In this subsection I would like to re-analyze
the characteristic features of the cavity model of MT in light of the recent experimental evidence on quantum coherent effects in protein antennae of marine cryptophyte algae ~\cite{algae}. This is essential, in the sense that the recent experiments gave us a feeling on the scale and the order of magnitude of the quantum coherent effects that appear to characterize biological complex systems at room temperatures.

First of all, let us remind the reader of the analogies between the cases of the photosynthetic algae and the MT, as far as their quantum aspects are concerned. In algae, one applies an external stimulus in the form of an external laser pulse, which excites quantum superpositions of electronic states in the protein-antennae pigments. There are coherent quantum oscillations and entanglement between pigments spatially separated at distances as long as 25 Angstr\"oms, at \emph{room temperatures}. The eventual environmental decoherence, induced by the rest of the algae complex, damps out such oscillations. The collapse of the pertinent wavefunctions occurs within 400 fs time intervals, which sets the time scale (\ref{algaedecoh}) as the characteristic time scale for decoherence of this system at ambient temperatures.

In photosynthetic systems like algae, the (quantum) oscillation periods are of order of a few hundreds of fs ~\cite{photocalc}. One may naively think that such short decoherence time scales have no relevance for any interesting biophysical process. However, the authors of \cite{algae}, and I tend to agree with them,  speculated that such a time period is sufficient for the bio-complex of algae to \emph{quantum compute} the most efficient, \emph{optimal}, way for energy and information transport across the pigments.

The experiments of ref.~\cite{algae}, as well as the relevant simulations of such bio systems~\cite{photocalc},
 constitute strong (experimental) evidence that quantum effects at ambient temperatures, playing a r\^ole in the system's functioning, are \emph{facts}, which, however, characterize \emph{some} but \emph{not all} photosynthetic biosystems. The cryptophyte algae
have the pigments co-valently bound with the rest of the complex, and this was argued in \cite{algae} to be the main reason why quantum coherence at ambient temperatures can occur, entangling parts of the complex at distances over 20 Angstr\"oms.

Tubulin dimers, on the other hand, have been theorized to be characterized by quantum superporsitions of their two conformations $\alpha$ and $\beta$. In this picture, the dimers constitute two-state quantum systems, oscillating between their two energy eigenstates with a frequency (\ref{frequency2}), corresponding to a period of $\sim 6000$~fs. For comparison, we mention that the characteristic period of quantum oscillations among the photosynthetic pigments in the calculated dynamics of \emph{Rhodopseudomonas acidophila}~\cite{photocalc} is about 500 fs (\emph{c.f}. fig.~\ref{fig:photocalc}), while in the photosynthetic algae studied in \cite{algae}, the period of the quantum oscillations is of order 60 fs. Presumably such quantum superpositions in tubulin can also be excited by external stimuli, such as applied (or self-generated~\cite{sataric,mn1}) electric and  magnetic fields, photon laser pulses \emph{etc.} When the quantum oscillating dimers find themselves bound in the walls of a MT, coherence has been argued to extend over macroscopic scales at distances of order $\mu$m, across the entire MT. The cavity model of MT has been invoked in \cite{mn1,mmn} as a plausible microscopic model underlying such properties. The unpaired charge of the tubulin dimer constitutes the two-state quantum system, and the rest of the tubulin protein complex, as well as the ordered water environment of the MT and its C-termini appendices~\cite{mt}, constitute the environment in this case.

If sufficient isolation from other electrostatic effects, such as Ions in MT~\cite{tegmark}, occurs in MT, then decoherence times as long as $10^{-6}-10^{-7}$~s can be achieved, (\ref{tdecohsoliton}).
As we have seen above, for this to happen~\cite{mn1} one needs strong electric dipole-dipole interactions, between tubulin protein dimers and ordered water molecules inside the thin cavity regions near the dimer walls.
This would provide an analogue to the strong co-valence binding of the pigments into the protein antennae of the cryptophyte algae~\cite{algae}.

In view of the fact that \emph{not all} photosynthetic systems exhibit such a strong binding, and that the latter has been linked~\cite{algae}
to the persistence of quantum coherence for relatively long times at ambient temperature, one is tempted to conjecture that such strong dipole-dipole interactions and cavity region realization may \emph{not} characterize
\emph{all} MT in \emph{all} biosystems. For instance, it would be interesting from the point of view of conscious perception, if only brain MT exhibited such strong cavity effects. Unfortunately the current experimental evidence towards this is nonexistent, and it is generally believed that there is a universal structure of MT, at least as far as the ordered water environment and the basic chemistry of tubulin dimers are concerned. But one cannot exclude surprises. Moreover, one should not dismiss the possibility that potentially important differences between \emph{in vitro} and \emph{in vivo} properties of biological matter may occur. The latter point of view is supported by the fact that living matter constitutes a non-equilibrium system, as we have also heard in the talk by E. del Giudice in this conference, and hence many concepts and properties we are accustomed to in physics, that pertain to equilibrium situations, may not characterize systems out of equilibrium.
Nevertheless, at present
such non universal behavior of MT constitutes a mere speculation, lacking any experimental confirmation or evidence.

In view of the \emph{experimental findings} of \cite{algae} on quantum decoherence times of order 400 fs, it would be interesting to examine
what range of the parameters of the MT cavity model of \cite{mn1} could lead to such short decoherence time scales. Of course,
MT and photosynthetic algae are completely different biological entities, and there is no \emph{a priori} reason why quantum effects, if any observed in MT, should be of the same order as in cryptophyte algae. Nevertheless, the question has a physical meaning, given in particular the fact that such short decoherence scales are of the same order as those induced by the neighboring ions in MT, (\ref{tegmark})~\cite{tegmark}.

In our model of \cite{mn1}, the main source of dissipation is assumed to come
from the loss of dipole quanta through the imperfect dimer walls.
From (\ref{convtime}), (\ref{lifetime}) we observe that an immediate six-order of magnitude reduction
in the decoherence time could be achieved by a corresponding reduction in the time $T_r$ over which the cavity MT dissipates its energy, which in turn would reflect on a much shorter life time of the alleged dipole quanta inside the cavity regions (\ref{lifetime}), which $T_r$ is identified with.
Such a reduction appears not natural within the super-radiance model of \cite{jibu}, used in our approach.
Indeed, the Rabbi coupling $\lambda$ that could conceivably change, even by several orders of magnitude, given the rough way its magnitude was estimated in our work so far, appears in (\ref{convtime}) inside a sinusoidal function, which is always less than unity. Hence, a change in the order of magnitude of $\lambda$ cannot reduce the decoherence time significantly.

The only parameters of the model  that could conceivably change and reduce (\ref{convtime}) by six orders of magnitude would be: (i) the density of ordered water molecules near the dimer walls $N_w/V$, which in our model had so far been assumed uniform inside the MT interior regions, and (ii) the number of coherent dipole quanta.
However, we find that changes of an unnatural magnitude in these two quantities would be required in order for (\ref{convtime}) to be reduced to the order of O(400) fs. It is hard to imagine, for instance, how we can arrange for $n \sim 10^6$ dipole quanta to be contained in a coherent cavity mode in the model.
The cavity model of \cite{mn1}, makes use of mesoscopic cavities, with $n$=O(10) at most, which is a natural number for dipole quanta inside the conjectured thin interior cavity regions of MT.

On the other hand, such short decoherence times in the quantum oscillations of the dimers, if observed, would indeed
point towards the fact that the cavity model of MT is in trouble, although quantum mechanics could still play an important r\^ole. For instance, as we have already discussed, short decoherence times of order (\ref{tegmark})
could be provided by the electrostatic effects of the  Ca$^{2+}$ and other ions~\cite{tegmark} in the C-termini and other neighborhoods of the dimers. In such a case, although the cavity model of MT would be probably invalidated, nonetheless we would still disagree with the conclusions of \cite{tegmark} that quantum effects play no r\^ole in efficient energy transfer in the (brain) cells. Following the speculations in  \cite{algae}, we may conjecture that such short times may be sufficient for the MT to \emph{quantum calculate} (through the quantum entanglement of (part of) its tubulin subunits) the \emph{optimal direction} along which energy and information would be most efficiently transported.
This would be a pretty important function. In this scenario, although energy would still be transported according to classical physics, nevertheless its optimum path would have been decided by a \emph{quantum computation}, lasting at most $10^{-13}$ s.

Another important aspect of the algae photosynthetic systems was the fact that they involved only a relatively small number of fundamental `units' (8) that undergo quantum superpositions and are entangled at room temperatures. On the other hand, the MT involve a rather large number of tubulin dimers, more than 100 in each protofilament.
It may actually be that in realistic situations \emph{not all}, but only a \emph{small subset} of relatively nearby tubulin dimers in a MT, separated by distances of up to 40 Angstr\"om, are actually ``coherently wired''. This could still be sufficient for the MT to \emph{quantum calculate} the \emph{optima}l direction for energy and signal transfer, as in algae protein antennae~\cite{algae}.

However, it may also be the case that the much longer --as compared to algae --  decoherence time scales in the cavity model of MT, can actually be explained, in a rather counterintuitive way, by the fact that a much larger number of fundamental ``units'' are coherently entangled in MT. Indeed, there are situations in physics where such a phenomenon occurs~\cite{mmn}. As mentioned in the introduction of the talk, in ref.~\cite{caes} the authors describe the
macroscopic  entanglement of two samples of Cs atoms at room
temperature. The entangling mechanism is a pulsed laser beam and
although the atoms are  far from cold or isolated from the
environment,  partial entanglement of bulk spin is unambiguously demonstrated
for  $10^{12}$ atoms for $\sim 0.5 ms$.  The system's resilience
to decoherence is in fact {\it facilitated} by the existence of a \emph{large
number of atoms}: even though atoms lose the proper  spin
orientation continuously, the {\it bulk} entanglement is not
immediately lost.  Quantum informatics, the science that deals
with  ways to encode, store and retrieve information written in
qubits, has to offer an alternative  way of interpreting the
surprising resilience of the Cs atoms by using the idea of ``\emph{redundancy}".
Simply  stated, information can be stored in such a way that the
logical (qu)bits correspond to many physical  (qu)bits and thus
are resistant to corruption of content.

Yet another way of looking
at this is given in the work of ref.~\cite{kielpinski}, where the authors have
demonstrated experimentally a \emph{decoherence-free quantum memory of one qubit} by encoding the qubit
into the ``decoherence-free subspace" (DFS) of a pair of trapped Berrilium
$^9Be^+$ ions. They achieved this by exploiting a
"safe-from-noise-area" of the Hilbert space for a {\it
superposition} of two basis states for the ions, thus encoding
the qubit in the superposition rather  than one of the basis
states. By doing this they achieved decoherence times on average  an
order of magnitude longer.

Both of the above works show that it
is possible to use DFS, error correction and high  redundancy to
both store information and to keep superpositions and
entanglements alive for biologically relevant times in macroscopic systems at high
temperature. Thus it nay  not be entirely inappropriate to
imagine that in biological {\it in vivo} regimes,
one has, under certain circumstances, such as the ones
specified above, similar entanglement of tubulin/MT arrangements~\cite{mmn}.

 I believe that very interesting future experiments can be done with MT, which could shed light on the above aspects of MT as quantum devices, which presently belong to the realm of \emph{science fiction}. Let me now discuss briefly some of such experiments.

\section{Experimental Tests of the MT Cavity Model: Past and Future \label{sec:3}}
\vspace{0.2cm}
I proceed now to discuss some experimental tests of the cavity model of MT presented in the previous section. I will concentrate on direct physical tests of the model, rather than expanding on its physiological or quantum information aspects. As I have already mentioned in the beginning of the talk, there are very interesting  experiments, for instance, showing consistency of the model as far as its predictions on the memory function of the brain are concerned~\cite{mershin,tus}, but I will not touch upon such aspects here. I am also not qualified to discuss experiments to study physiological/biological properties of MT and/or the tubulin dimers in general.

Some of the experiments I will describe below, require for their interpretation detailed theoretical knowledge on the structure of the tubulin dimers and MT, especially as far as their electrical properties are concerned.
Computer molecular simulations for the permanent electric dipole moments of the tubulin, to be discussed briefly below, can be found in \cite{tusedm,mershinedm} and are based on the structure of the tubulin dimer provided by electron-microscope studies at 3.5 Angstr\"om resolution in ref.~\cite{downing}. For detailed recent molecular dynamics simulations on the structure of the MT, which may also be useful in future studies and tests of properties of MT,
including our cavity model, I refer the reader to the interesting works of ref.~\cite{recentMTstr}. However, for the past experiments I will discuss below, we used the simulations and tubulin structure data available up to the year of the experiments (2005).

\subsection{Measurement of Electric Dipole Moments and Ferroelectric Properties Tests}
\vspace{0.2cm}
An essential aspect of the model is its ferroelectric features at room temperatures, namely the induction of a permanent electric dipole moment that remains after switching off an externally applied electric field.
We have performed such experiments with porcine brain MT in 2005, in collaboration with the molecular biology group of E. Unger~\cite{stracke}.
\begin{figure}[ht]
\begin{center}
  \includegraphics[width=4cm, angle=-90]{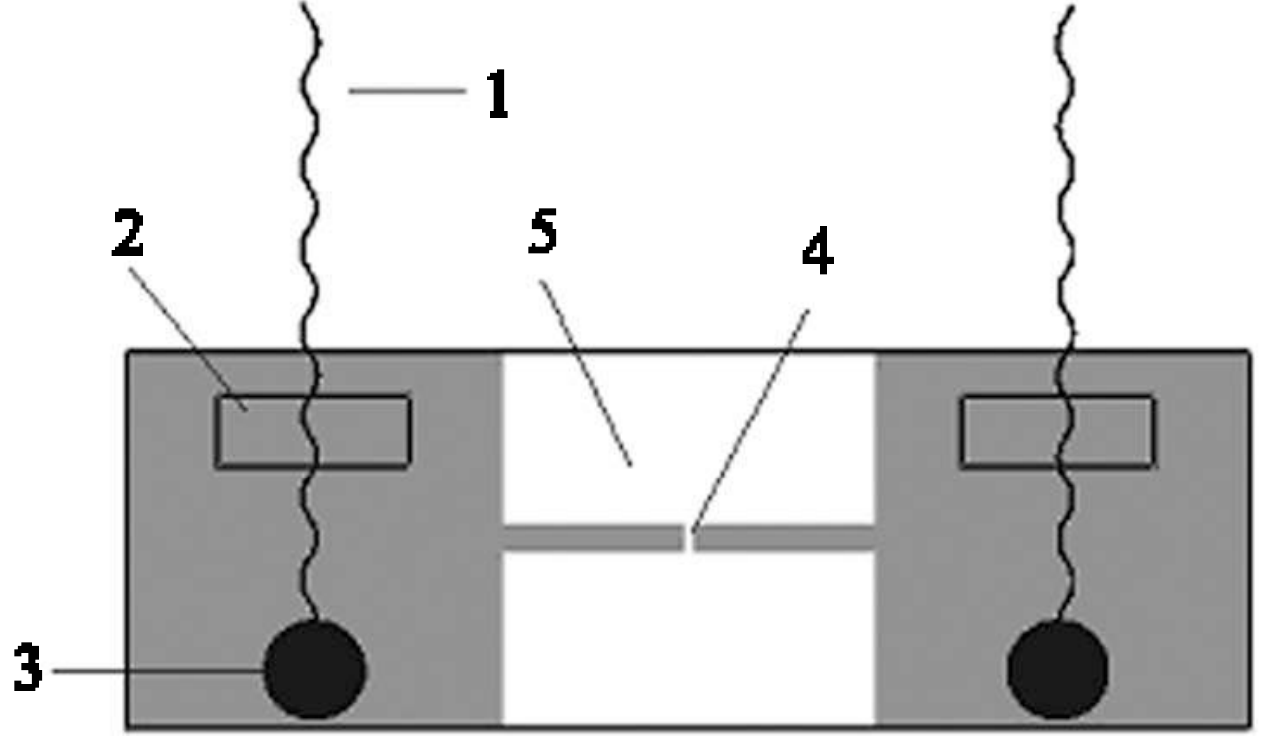}\hfill \includegraphics[width=4cm, angle=-90]{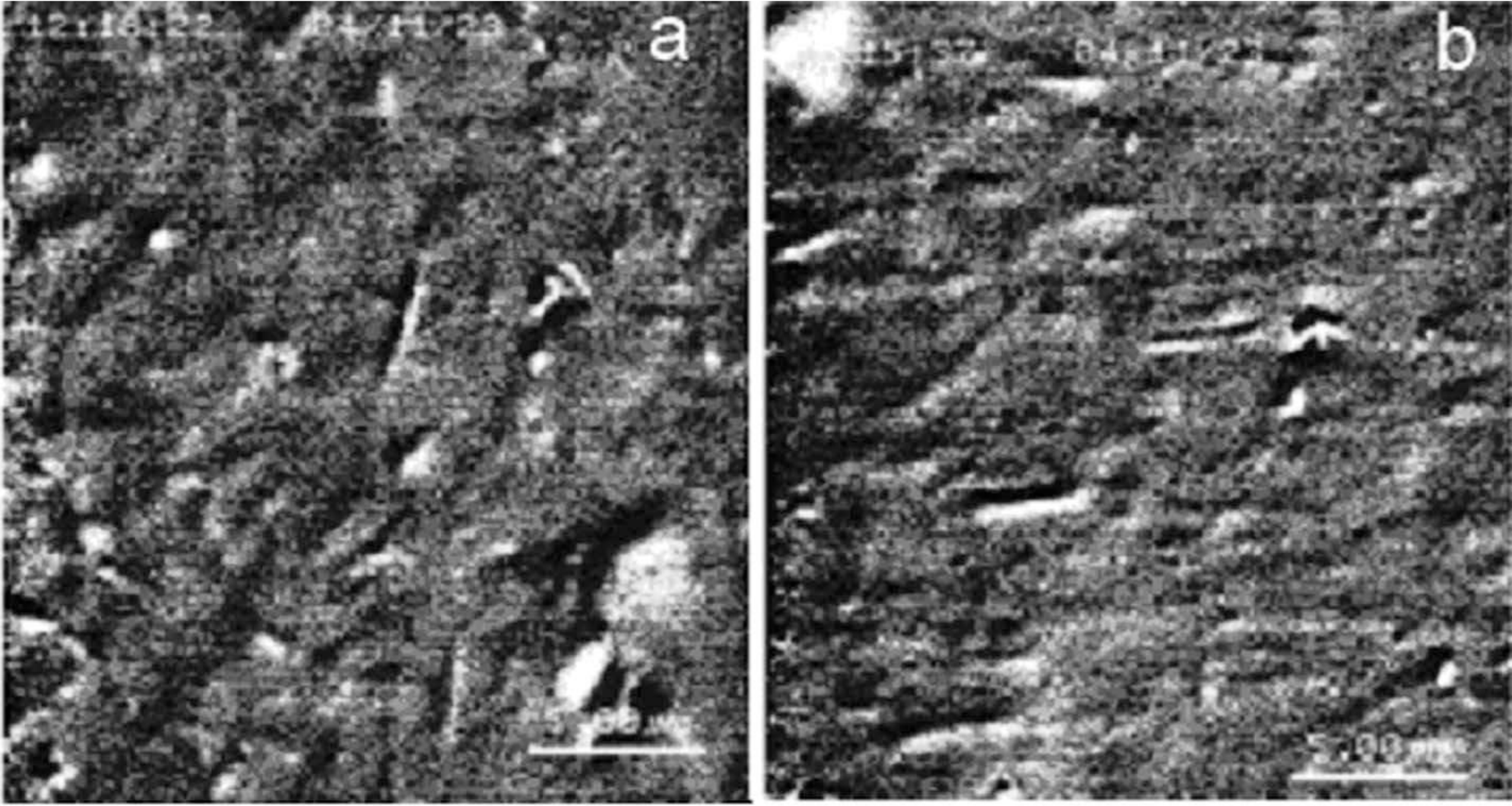}
\end{center}
\caption{\emph{\textbf{Left Picture}}: Experimental arrangement for measuring the electron dipole moment of MT in suspension. The picture shows the electrode system used for the measurements in alternating electric fields. The gray areas denote gold layers. 1: thin, flexible wires for the application of the voltage; 2: strips of adhesive tape as pull relief; 3: drops of conductive silver for contacting the wires with the gold; 4: electrode gap filled with the sample of porcine brain MT; 5: gold-free glass areas.
\emph{\textbf{Middle and Right Pictures:}} \textbf{a.(Middle:)} Random positions of MT in the absence of electric field. This state is again realized after switching off the field.
\textbf{b. (Right):} Alignment of MT in the direction of the externally applied alternating electric field
210,000 V/m and 2 MHz frequency. The visualization of MT has been achieved by video-enhanced DIC Microscopy
(Pictures taken from ref.~\cite{stracke}).}
\label{fig:unger}
\end{figure}
We have examined the effects of both constant (up to $2 \times  103$ V/m) and high-frequency alternating fields
(up to $2.1 \times  105$ V/m, with frequencies from 200 kHz to 2 MHz) on suspended porcine microtubules.
At pH 6.8 and 120 mM ionic strength, constant fields cause a motion of microtubules
toward the anode (\emph{c.f}. fig.~\ref{fig:unger}).

The electrophoretic mobility amounts to $2.6 \times 10^{-4}~$ cm$^{2}$/V\,s,
reflecting a negative net charge of approximately 0.2 elementary charges per tubulin
dimer. The moving microtubules are randomly space oriented. Alternating high-frequency
fields induce electric dipoles and align the microtubules parallel to the
field direction. By determining the angular velocity of the turning microtubules, we
estimate a dipole moment for the MT roughly
\be\label{dipoleMTexp}
p_{\rm MT} = 34,000 ~{\rm Debye}
\ee
at $2.1 \times 105$ V/m and 2MHz frequency.
By
comparing the potential energy of the dipole in the applied field with the thermal
energy of microtubules, we obtained a minimum value of 6,000 Debye, necessary
for efficient alignment.

Unfortunately \emph{no evidence} for permanent electric dipole has been found at ambient temperatures, where the experiment has been performed. Thus ferroelectricity has \emph{not} been confirmed as yet.
 In what follows, I will attempt a comparison of these results with theoretical estimates and seek possible explanations for our inability to observe a permanent electric dipole moment, other than the straightforward dismissal of any ferroelectric properties of MT.

Computer molecular simulations of the permanent electric dipole moment of the tubulin dimers~\cite{tusedm}, taking into account their detailed structure~\cite{downing}, have shown that the bulk of it is directed in a direction perpendicular to the protofilament axis ($x$-axis) of the MT, and only about a fifth of the total electric dipole moment is along the $x$-axis:
\be\label{edmdimer}
p_x = 337 ~{\rm Debye}~, \qquad p_y = -1669 ~{\rm Debye}~, \qquad p_z = 198 ~{\rm Debye}~.
\ee
Taking into account the screening effect of water
(whose dielectric constant may be as high as 80) on the dimer charges, yields
the following suppressed estimate for the (permanent) dimer dipole moment~\cite{mershinedm}
\be\label{dimerwater}
p_{\rm dimer} = 90~{\rm  Debye}~.
\ee
The unpaired charges,that is, the net charges inside the hydrophobic pockets of
the dimers, that appear isolated from their environment, may lead to even further
suppression , that is, one may take as an
order of magnitude~\cite{stracke}
\be
p_{\rm dimer} = 15~{\rm  Debye}~.
\ee
One may speculate that the other components perpendicular to the microtubule
axis will be neutralized in the cylindrical microtubular geometry and screened by
the environment, thereby leaving the $p_x$ component as the dominant contribution
to the total dipole moment of the microtubule. The reader is reminded at this point that such an assumption also characterizes the simplified ferroelectric one-dimensional lattice model for MT, discussed in \cite{sataric} and adopted in our studies in \cite{mn1}.

If this is the case, then taking into
account that in moderately long microtubules of an average length L = 3.5 $\mu $m
with 12 protofilaments each (as used in the experiments of \cite{stracke}), there are about ${\cal N}$ = 5280 tubulin dimers of average length 8 nm each, one would arrive
at the most optimistic estimate for the total dipole moment (all dimer dipoles
contributing equally to the x-direction) in the range
\be\label{estimateMT}
p_{\rm total} = {\cal N} \, p_{\rm dimer} = 79,200~{\rm  Debye}~.
\ee
This is likely to be further suppressed if details on the water and other
environments and geometry are properly taken into account.

Such a suppression of the total dipole moment may provide an explanation for
our inability to observe an alignment of microtubules in constant electric fields up to
$2 \times 103$~ V/m. As discussed in \cite{stracke}, estimates of the interaction energy of the supposed permanent dipoles in the electric fields show that this energy is too small to overcome the influence of
the thermal energy. Safety requirements prevented the use of higher intensity constant fields in our experiments.
On the other hand, as we already mentioned, the application of alternating fields with intensities as high as
$2 \times  105$~ V/m, at 2~MHz frequency, induce
alignment of the MT along the direction of the field (\emph{c.f.} fig.~\ref{fig:unger}), from which we estimated the dipole moment (\ref{dipoleMTexp}).

The results of this experiment, therefore, probably imply that
a possible permanent part of the dipole moment
does not play a role for the orientation at high frequencies because the dipole
cannot follow the changes of the field. Only the induced part of the dipole is
responsible for the orientation, because the torque does not depend on the field
direction. Comparing the potential energy of the dipole in the applied field with the
thermal energy of microtubules, shows, as mentioned above, that a minimum value of the dipole moment
is necessary for a successful orientation.

Thus the non-observation of ferroelectric properties in our experiment does not falsify the ferroelectric-ferrodistortive models of MT~\cite{sataric} on which the cavity approach is based~\cite{mn1,mmn}.
Further, refined, experiments in this direction are certainly due.

\subsection{Quantum-Optics-Inspired Experiments for the Cavity Model of MT: are they feasible?}
\vspace{0.2cm}
One of the most important ingredients in the approach of \cite{mn1} was the r\^ole of the coherent cavity modes of the electric dipole quanta. Their (Rabi) dipole coupling with the dipoles of the tubulin dimers provides the necessary ``friction'' environment, as we discussed above, which is responsible for the soliton (coherent) states that transfer energy across the MT in a dissipation-free manner.
One of the decisive tests of the MT cavity model, therefore, would be to observe such coherent modes/Rabi couplings, adapting properly the pertinent quantum optics/cavity electrodynamics experiments~\cite{rabiexp,haroche} in the MT situation. For completeness, below I will discuss these experiments, with a view to see whether there is a possibility of applying them to the biological systems at hand.
\begin{figure}[ht]
\begin{center}
  \includegraphics[width=4cm, angle=-90]{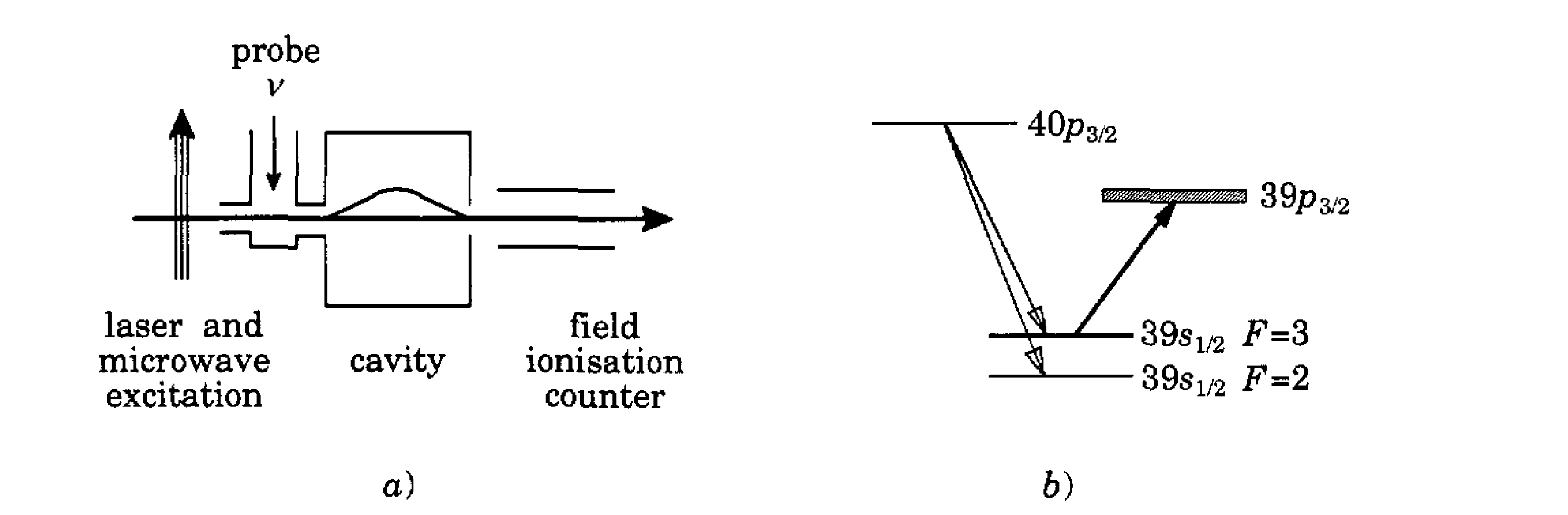}\hfill \includegraphics[width=4cm, angle=-90]{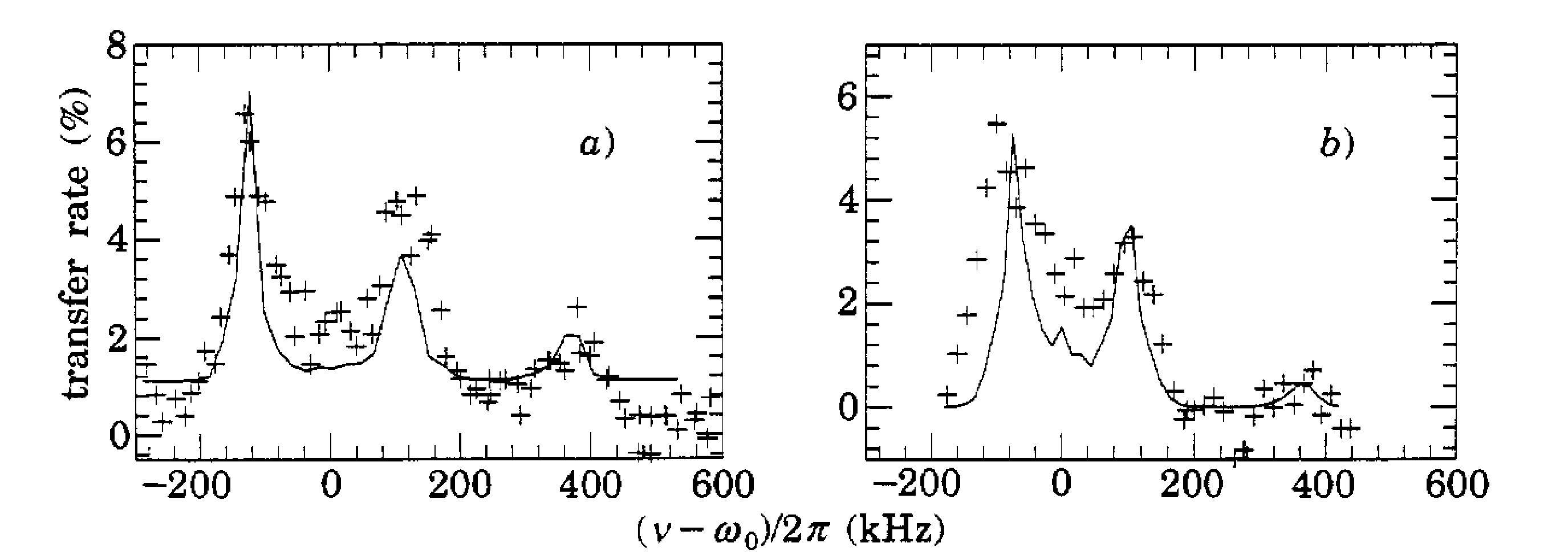}
\end{center}
\caption{\emph{\textbf{Upper Figures}}: \textbf{a)} Experimental set up  for the measurement of the Vacuum Field Rabi splitting in atomic physics. The atomic beam is represented by a horizontal arrow. The atoms are prepared in some initial configurations (39s$_{1/2}$ with hyperfine two-state splitting $F=2, 3$, indicated by arrows pointing downwards) by laser and microwave excitation. The probing electric field (with a frequency $\nu$ and sine arch time variation of its amplitude, indicated in the figure by a continuous curve inside the cavity region) is injected into the cavity through the same wave guides as the atoms. \textbf{b.) } The cavity mode couples the 39s$_{1/2}$ F=3 hyperfine level to the 39p$_{3/2}$ levels (their hyperfine structure remains unresolved in the conditions of the experiment). This is indicated by an arrow pointing upwards. \emph{\textbf{Lower Figures:}} The observed Rabi spectra, corresponding to the following number of atoms in the cavity: \textbf{a.}) N = 10, \textbf{b.}) N = 5. Crosses indicated experimental points, continuous lines are theoretical simulations. $\omega_0/2\pi =$~68.38 GHz  is the frequency of the 39$s_{1/2}$ F=3 $\rightarrow$ 39$p_{3/2}$ transition. The transition 39$s_{1/2}$ F=2 $\rightarrow$ 39$p_{3/2}$ has a shifted frequency
$(\omega_0 + \Delta)/2\pi$, $\Delta/2\pi = 320 $~KHz \,(Pictures and results taken from ref.~\cite{rabiexp}).}
\label{fig:rabi}
\end{figure}

In cavity quantum electrodynamics, the easiest way to observe the Rabi splitting of atoms~\cite{rabiexp}, discussed briefly in subsection \ref{sec:mtcavity} above,  (\ref{rabiabs}), (\ref{rabisplitting}), is through the experimental set up indicated in fig.~\ref{fig:rabi}.
Some two-state Rydberg atoms (or, more generally (\emph{c.f}. fig.~\ref{fig:rabi}), atoms properly prepared so as to resemble two-state systems) are injected through a wave guide into a cylindrical superconducting cavity, which in the experiments of \cite{rabiexp} is cooled down to 1.7 K. Before entering the cavity the atoms are prepared in some configuration of the hyperfine splitted 39$s_{1/2}$ F=2, and F=3 states. The cavity mode couples the
39$s_{1/2}$ F= 3 state to the 39$p_{3/2}$ state. As a result of this coupling, upon the application of the external electric field of frequency $\nu$, the cavity mode does not resonate at the bare atom frequency but exhibits instead
 two peaks in the corresponding absorption spectrum, as indicated in the lower part of fig.~\ref{fig:rabi}.
 From these peaks one clearly can compute the Rabi coupling, using (\ref{rabiabs}) or (\ref{rabisplitting}).

 However, experimentally (\emph{c.f.} fig.~\ref{fig:rabi}) there are more features and structures than the
 two symmetric absorption peaks theory predicts: the two observed Rabi main peaks appear asymmetric and
 there is a much weaker third peak at frequencies around $(\omega_0 + \Delta)/2\pi$.
 The latter feature is clearly associated with the non-resonant coupling of the 39$s_{1/2}$ F=2 atoms to the cavity
 modes, since $(\omega_0 + \Delta)/2\pi$ is the frequency of the 39$s_{1/2}$ F=3 $\rightarrow$ 39$p_{3/2}$ transition
 (\emph{c.f}. fig.~\ref{fig:rabi}). The former are associated with the motion of the atoms, as well as the fluctuations in the atom number inside the cavity. Numerical simulations of such effects confirm the experimental results and thus provide convincing explanations of the Rabi-splitting phenomenon in realistic systems~\cite{rabiexp}.

 Such measurements are then used for the classification of atoms that are employed in quantum-optics experimental demonstrations of environmentally-induced decoherence effects in atomic physics~\cite{harochedecoh}. In these experiments, two-state atoms in a quantum superporition are sent through isolated cavities filled by microwaves. The corresponding Rabi couplings
 of the atoms to the cavity coherent modes cause a shift in the phase of the microwave field, by different amounts.
  The experiment involves Rydberg atoms that interact one at a time with the few photon coherent modes ($O(1-10)$), trapped in the cavity. In this way, the field in the cavity is also put in a superposition of two states. The exchange of energy of the field with its environment, and the loss of photon coherent cavity modes through the (imperfect) cavity walls, imply decoherence and therefore the eventual collapse of the field superposition into a single definite state. The authors of \cite{harochedecoh} have observed experimentally this decoherence, while it unfolded,  via the study of correlations between the energy levels of pairs of atoms sent through the cavity  with various time delays between the atoms.

  In the case of the cavity model of MT, there are formal similarities between the quantum superpositions of the two energy levels of the two-state atoms used in the above-described quantum optics experiments and those of the two conformations of the tubulin dimers.
  However, from the technical point of view, the situation is much more complicated.
  Unlike atoms, the MT complexes contain many entities in their environment, exhibiting motion, vibrations (due to (room) temperature effects) \emph{etc}, which complicate simple tests in search of Rabi-couplings
  between the dimer excitations, playing the r\^ole of the atoms, and the electric-dipole quanta inside the cavity regions of the MT, playing the r\^ole of the
  quantized cavity field modes in the atomic physics experiments.
  Such couplings would lead~\cite{mn1} to frequency splittings, with the characteristic peaks (\emph{c.f}. fig.~\ref{fig:rabi})
  (\ref{rabiabs}), (\ref{rabisplitting}), in the appropriate absorption spectra of MT
   (see discussion in subsection \ref{sec:mtcavity}).

  Unfortunately, I am not qualified to discuss in detail the feasibility of such experiments. I can only make speculations which, from an experimental point of view, may not be realizable in practice. Nevertheless, as this is a talk, I am free to speculate, so here are my thoughts/suggestions on such experiments:
  one needs, first of all, to have isolated microtubules, probably in suspension. One should apply laser fields through them and study the absorption spectrum. Since the conformational changes of the tubulin dimers in MT are expected
   theoretically~\cite{mn1} to be in the thousand GHz region (\ref{frequency2}), one should arrange for the externally applied fields to have frequencies near such values. According to the model of cavity MT, reviewed above, the characteristic coherent modes of the cavity regions of the MT, excited by the field, are almost in resonance with the dimer oscillations, and hence standard Rabi-splitting phenomenology (\ref{rabisplitting}) should be expected, if the model correctly describes nature.

    The induced Rabi splitting in the frequencies of the absorption spectra of MT, if observed, would then constitute compelling evidence for the existence of \emph{both} isolated cavity regions in MT interiors and coherent modes inside such regions, which in the model of \cite{mn1} would be the dipole quanta of the water molecules, suggested in \cite{vitiello}. It goes without saying, that in view of the complicated MT structure, and the associated motions and vibrations of the various biological entities in the tubulin protein or the C-termini appendices, there would be non resonant couplings of atoms and ions in the dimers to the cavity electron dipole quanta, in addition to the simple dipole Rabi couplings examined so far. These would complicate the resulting spectra, however one should still expect to see pronounced Rabi peaks, in analogy with the atomic physics experiments of fig.~\ref{fig:rabi}.

    Unfortunately,
    observations of such effects alone (even if they are realized in nature) would not constitute a proof of the quantized nature of the dimer excitations of the MT. Although the latter is a plausibility, and according to such interpretations, the (conjectural) thin interior cavity regions of MT would entail vacuum fluctuations of the ordered-water dipole quanta that would split the resonance line of the dimers by an amount proportional to the
    collective dimer-water-cavity coupling (a sort of induced dynamical Stark Effect on the quantum superpositions of the dimer states), alternative explanations of the Rabi splitting phenomenon~\cite{rabiexp} exist. The latter point towards
    an interpretation of the phenomenon as a consequence of the fact that the MT dimer \emph{medium} behaves as a \emph{refractive} one with a \emph{classical} \emph{complex} (\emph{i.e} containing imaginary parts) index of refraction that splits the cavity mode into two components.

    To demonstrate, therefore, unambiguously the existence of quantum coherent effects in MT one needs to explicitly observe experimentally
the quantum oscillations of the dimers and measure their environmentally induced decoherence.
A straightforward extension of the  atomic physics experiments, observing quantum decoherence in electrodynamical
cavities~\cite{harochedecoh} may, unfortunately, not be feasible in the case of MT. Indeed, in the case of
the atomic physics experiments of ref.~\cite{harochedecoh}, a Rydberg atom beam in a quantum superposition is sent through a cavity, \emph{exits} from it and is eventually counted in one or the other Rydberg state by appropriate ionizing detectors, so that only decoherence of the atom-cavity system (``atom plus measuring apparatus'') is measured in the experiment. In contrast, in the case of MT, the ``cavity'' regions are attached to the ``atoms'', being part of the MT structure. One cannot separate the quantum oscillations of the dimers from the rest of the MT and its ordered water interiors. Thus, the methods of \cite{harochedecoh}, that could in principle measure quantum decoherence of the field-dimer (and hence of the entire MT) system, as the latter unfolds, appear inapplicable.

On the other hand, it seems to me that Photon echo absorption data, like the ones in \emph{Cryptophyte Algae}~\cite{algae}, when appropriately adapted to the much more complex case of MT, might be a way forward, in order to observe the decoherence of the coupled ``field-dimer'' system inside the MT. Presently I do not know whether such measurements are feasible in  the near future, nevertheless, I find the prospect of performing such experiments very exciting, and I am sure there are ways one can proceed along these lines in the near future.

\section{Conclusions and Outlook \label{sec:4}}
\vspace{0.2cm}
The exciting experimental developments in the light-harvesting \emph{Cryptophyte Algae} provided compelling evidence on an important r\^ole of quantum effects in biological systems at room temperature. Specifically, on exciting by photon pulses certain dimer pigments (DVB) of the photosynthetic protein antennae of the algae into a quantum superposition of appropriate electronic states, one observes experimentally, by means of two-dimensional photon echo (absorption) data~\cite{algae}, the quantum oscillations between the two electronic states of the DVB dimers, as well the quantum entanglement of the DVB molecules with the other pigments, at distances of order 20 Angstrom away. In this way, the entanglement is responsible for ``coherently wiring'' pigment molecules across the entire protein antennae.
The experiment has been performed at ambient temperatures (294 K).

The situation is reminiscent of coupled oscillators through extended springs. Such action at a distance is the result of quantum correlations between the quantum states of the pigments. The eventual decoherence of the relevant oscillations, induced by the complex environment of the protein antennae, has been observed to last for about 400 fs.
Although this time scale is relatively short, nevertheless, the authors of ref.~\cite{algae} have argued that it may be sufficiently long for the protein antenna to \emph{quantum calculate} in which direction energy and information would be transported more efficiently. Thus the observed coherent `wiring' across the entire antenna complex is thereby linked with energy transfer optimization in photosynthetic algae.

Not all photosynthetic proteins exhibit such a behavior, which in \cite{algae} has been attributed to the fact that the pigments are covalently bound in the protein complex, unlike the case of other light-harvesting biosystems.
This strong bounding is probably responsible for the rather long (from the point of view of information transfer optimization) decoherence time scales at room temperatures.

Such exciting developments brought back to life, at least in the mind of the speaker (!), a rather old model of biological microtubules (MT) viewed as quantum electrodynamical cavities, developed in \cite{mn1} and elaborated further in \cite{mmn}. In this model, quantum coherent oscillators in the dimer walls was the result of environmental entanglement of the tubulin protein dimers (viewed as a two state quantum system, oscillating between the two conformational states, $\alpha$ and $\beta$ - tubulin) with quantum dipole quanta inside thin interior regions of the
MT, extending from the dimer walls inwards, for a few Angstr\"oms, which are full of ordered water. These regions have been argued to play the r\^ole of thermally isolated cavity regions, as a result of strong dipole-dipole interactions between the electric dipoles of the dimers and those of the dipole quanta in the ordered water. The latter arise as a result of the interaction of water molecules with quantized electromagnetic fields inside the cavity regions.

Electromagnetic fields in MT can be either externally applied or self generated. In this talk, it has been argued that strong experimental evidence in favor of the existence of dipole quanta in the interior of MT would be the observation of characteristic Rabi splitting in the photon absorption frequency spectra of tubulin dimers, under the influence of
externally applied electromagnetic fields of definite frequency.

The dipole-dipole water/dimer coupling is held responsible~\cite{mn1} for inducing macroscopic quantum coherence across the entire MT, of some $\mu $m=10$^{-6}$~m long, which would manifest itself as a dimer-dipole kink soliton, responsible for dissipation-free energy and signal transfer. It has been estimated that, within some natural range of the parameters of the model of \cite{mn1}, the decoherence time scale, within which such solitonic states collapse to a classical state, could be as long as
\be
t_{\rm MT~decoh} \mathcal{O}\left(10^{-6} - 10^{-7}\right)~{\rm s}~.
\ee
under the assumption that the dominant source of decoherence
is the loss of dipole quanta through the imperfect cavity walls.
Such long decoherence times may be achieved if there is ferroelectric behaviour of MT complexes at room temperature. They are argued to be sufficiently long for dissipation-free energy and signal transfer by the kink soliton  along the MT protofilaments.
\begin{figure}[t]
\begin{center}
  \includegraphics[width=7cm, angle=-90]{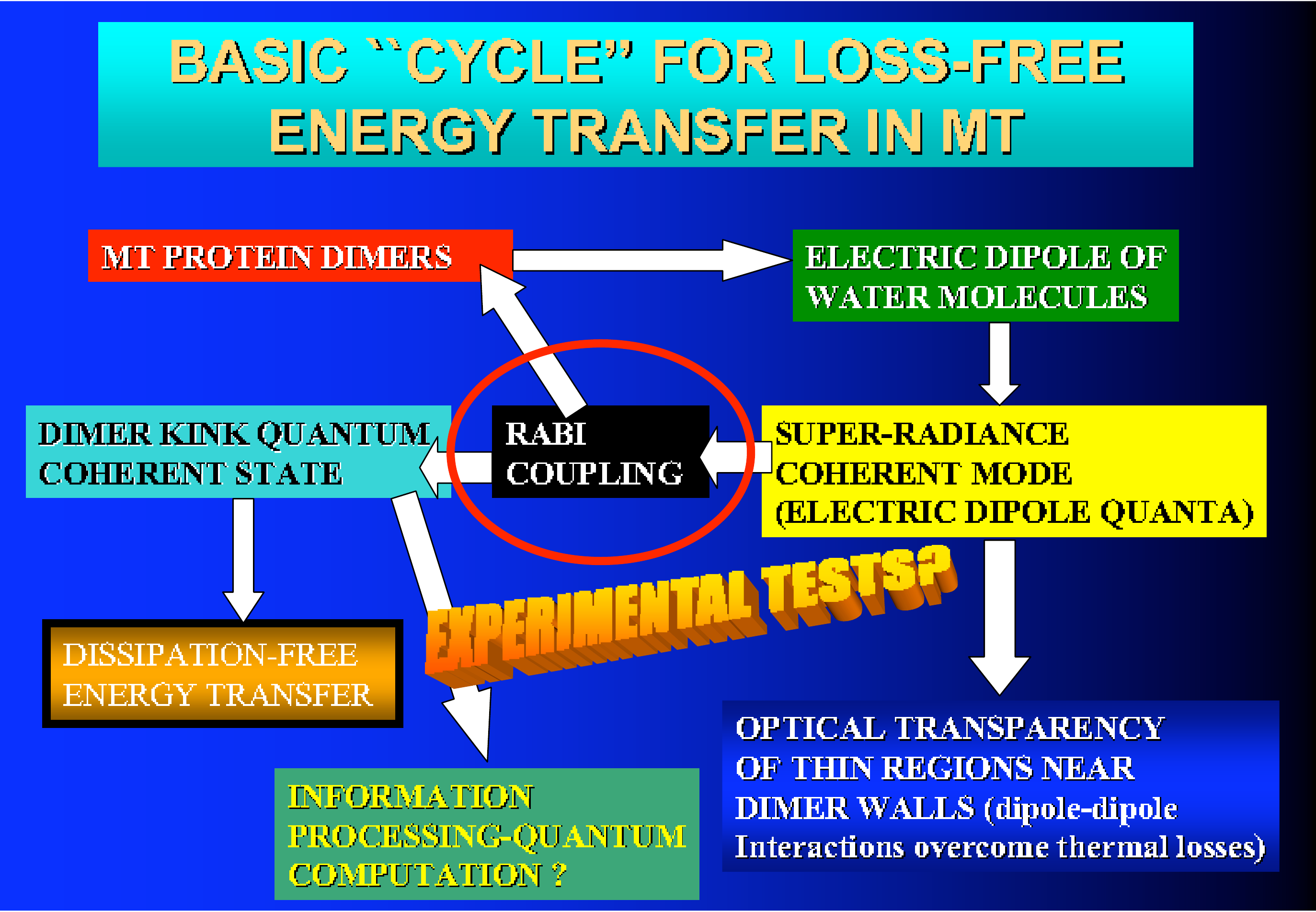}
\end{center}
\caption{Suggested basic cycle, summarizing functioning and properties of the quantum cavity model for Microtubules~\cite{mn1}, and its r\^ole in dissipation-free energy and information transfer.}
\label{fig:cycle}
\end{figure}
Our mechanism for the emergence of coherent states
in MT arrangements therefore inside the cell can be summarized
by the `basic cycle' shown in figure \ref{fig:cycle}.

Much shorter decoherence times of the quantum excitations of the dimers are expected if the ferroelectric behaviour and the strong dipole-dipole water/dimer coupling do not exist in Nature. In particuler, as explained in \cite{tegmark}, decoherence times of order of a few hundreds of fs may be the result of Ca$^{2\,+}$-ion induced decoherence in MT.
Although, such generic estimates have been argued in \cite{tegmark} to eliminate any essential r\^ole of quantum physics in biological MT, in particular brain MT, nevertheless, in view of the findings of \cite{algae}, such conclusions should be revised.

Decoherence times, of order of a few hundreds of fs, \emph{may be} sufficiently long for optimization of the direction along which energy and information transfer in MT would be more efficient, even if the latter occurs by means of classical physics (mechanics) laws of transport rather than quantum.
Thus, even if the cavity model of MT turns out \emph{not} to provide the correct description of natural MT, some sort of quantum computation in energy and signal optimization could still take place inside the  biological cells, in the way conjectured above.

All such features would leave the realm of \emph{science fiction} and become \emph{realistic possibilities} or even realities, once some experimental evidence on the basic properties of models entailing quantum dynamics of MT emerges.
The latter can come in many ways, from a direct observation of quantum oscillations of dimers in an MT, using photon echo data as in \cite{algae}, assuming that such a technique is feasible in MT,
  and indirect verification of the existence of cavity regions in their interior, through the observation of Rabi-splitting peaks in appropriate absorption spectra,
to an explicit demonstration of ferroelectric properties of MT at room temperatures, through the measurement of permanent dipole moments and other phenomena characteristic of ferroelectricity~\cite{zioutas}.
Unfortunately, at present there is no such experimental evidence.
However, the situation can easily change in the near future.

I would like to close this talk by stating once more something that physicists tend to forget, when experimenting with
living matter. \emph{In vitro} (Laboratory) situations may be entirely different from actual \emph{in vivo} situations. Living organisms are non equilibrium systems, as emphasized in this conference by E. del Giudice, hence,  their behaviour may be entirely different from dead \emph{in vitro} samples in suspension.
It is then possible that some quantum aspects of \emph{in vivo} Microtubules are not easily (if at all) revealed  in the laboratory. More sophisticated than usual techniques might be needed to bring forward such aspects.

I sincerely believe that the emergence of quantum phenomena in light-harvesting algae will open new and exciting avenues for research in this direction, and, if such coherent phenomena prove to characterize also
the fundamental building blocks of the cell, the Microtubules, this may eventually lead to establishing a possibly important r\^ole of quantum physics on basic functions of the cell.
Time will probably tell whether all these constitute \emph{science fiction} or \emph{facts}...

\section*{Acknowledgements}
\vspace{0.2cm}
I thank the organizers of the DICE 2010 Workshop (Castello Pasquini, Castiglioncello (Italy), September 11-18 2010) for the invitation and the opportunity to talk in this interesting and thought stimulating conference. I also acknowledge informative discussions on molecular modelling with C. Molteni. The work of N.E.M. is partially supported by the European Union through the FP6 Marie
Curie Research and Training Network \emph{UniverseNet} (MRTN-CT-2006-035863).

\end{document}